\documentclass[10pt,twocolumn,english,conference]{IEEEtran}
\usepackage[T1]{fontenc}
\usepackage[latin9]{inputenc}
\usepackage[a4paper]{geometry}
\usepackage[active]{srcltx}
\usepackage{color}
\usepackage{array}
\usepackage{booktabs}
\usepackage{textcomp}
\usepackage{multirow}
\usepackage{amsmath}
\usepackage{amssymb}
\usepackage{graphicx}
\usepackage{fancyhdr}

\makeatletter

\providecommand{\tabularnewline}{\\}

\usepackage{times}\usepackage{epsfig}\usepackage{amsfonts}

\makeatother

\usepackage{babel}

\makeatother

\usepackage{babel}

\makeatother

\usepackage{babel}

\usepackage{url}

\usepackage{graphicx}
\usepackage{caption}
\usepackage{cite}

\makeatother

\usepackage{babel}
\begin{document}
\title{A Survey on Radio Frequency Identification as a Scalable Technology
to Face Pandemics}
\author{{Giulio M. Bianco}, {Cecilia Occhiuzzi}, {Nicoletta Panunzio},
and {Gaetano Marrocco}\\
\\
}
\maketitle
\begin{abstract}
The COVID-19 pandemic drastically changed our way of living. To minimize
life losses, multi-level strategies requiring collective efforts were
adopted while waiting for the vaccines\textquoteright{} rollout. The
management of such complex processes has taken benefit from the rising
framework of the Internet of Things (IoT), and particularly the Radiofrequency
Identification (RFID) since it is probably the most suitable approach
to both the micro (user) and the macro (processes) scale. Hence, a
single infrastructure can support both the logistic and monitoring
issues related to the war against a pandemic. Based on the COVID-19
experience, this paper is a survey on how state-of-the-art RFID systems
can be employed in facing future pandemic outbreaks. The three pillars
of the contrast of the pandemic are addressed: 1) use of Personal
Protective Equipment (PPE), 2) access control and social distancing,
and 3) early detection of symptoms. For each class, the envisaged
RFID devices and procedures are discussed based on the available technology
and the current worldwide research. This survey that RFID could generate
an extraordinary amount of data so that complementary paradigms of
Edge Computing and Artificial intelligence can be tightly integrated
to extract profiles and identify anomalous events in compliance with
privacy and security.
\end{abstract}


\section{Introduction}

\captionsetup[figure]{labelfont={default},labelformat={default},labelsep=period,name={Fig.}} \captionsetup[table]{labelfont={default},labelformat={default},labelsep=newline,name={TABLE},textfont={sc},justification=centering}\let\thefootnote\relax\footnotetext{Authors are with the Pervasive Electromagnetics Lab, University of Rome Tor Vergata, Rome, Italy. www.pwevasive.ing.uniroma2.it. Corresponding author: gaetano.marrocco@uniroma2.it}
\let\thefootnote\relax\footnotetext{This is the author's version of an article that has been accepted for publication in \textsl{IEEE Journal of Radio Frequency Identification}. Changes were made to this version by the publisher prior to publication. The final version of record is available at \color{blue}{http://dx.doi.org/10.1109/JRFID.2021.3117764}}

COVID-19 pandemic abruptly changed the way we live, starting from
social interactions and healthcare, up to commerce, transportation
and entertainment. The high spreading capabilities (even by asymptomatic
individuals \cite{Arons20}) and the severe impact on public health
\cite{Hospitalization} forced the Governments to adopt urgent restrictive
countermeasures \cite{Qian2020}. Although at the time of writing
the emergency is mitigating, the earned experience so far has taught
us that at least two consecutive phases must be promptly managed to
proficiently face the future pandemics and allow a rapid recover of
activities and daily life \cite{Whitelaw20}: (\emph{i}) social distancing
with home confinement and/or lockdown until the availability of vaccines,
(\emph{ii})\emph{ }technological support to mitigate the infection
risk during reopenings, to manage COVID-19 patients in hospitals and
at home, and to handle the logistics issues of the distribution of
vaccines \cite{WHOvaccines}.

The pillars of countermeasures against a respiratory infectious pandemic
can be identified as:

(\emph{i}) the usage of Personal Protective Equipment (PPE), like
facemasks and hygiene products, to reduce the spread of infecting
particles;

(\emph{ii}) social distancing and access control to avoid, or at least
reduce, possible channels of infections among people, and contact
tracing to identify recent interactions with infected people;

(\emph{iii}) early detection of symptoms (e.g. fever, blood saturation,
irregular breath, cough, in synergy with Nasopharyngeal/Oropharyngeal
Swabs and serologic tests).

Furthermore, due to the scientific and clinical evidence of the presence
of infectious droplets floating in the air for a long time, especially
in indoor environments \cite{vanDoremalen20}, the continuous verification
of air quality and ventilation is recommended to reduce the risk of
inhalation and of infection \cite{deMan20}.
\begin{figure*}[h]
\begin{centering}
\includegraphics[width=18.1cm]{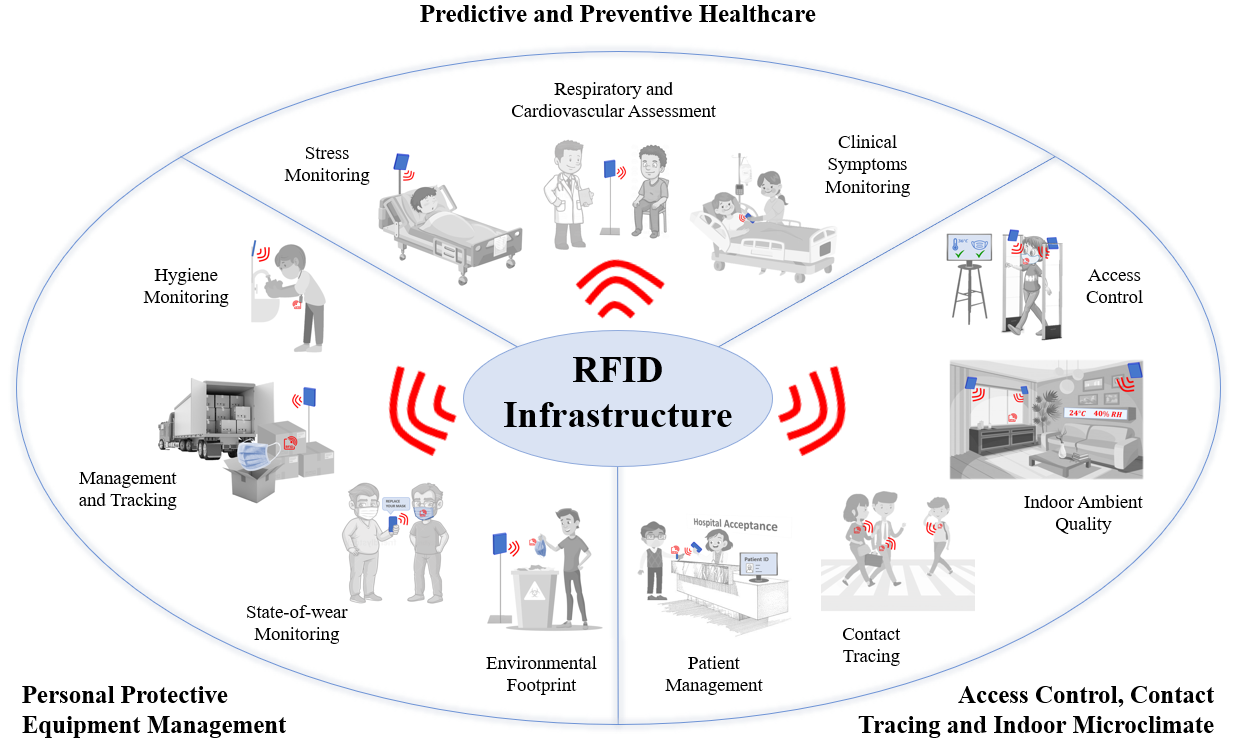}
\par\end{centering}
\caption{Synopsis of applications of RFID technology to pandemic countermeasures.
\label{fig:RFID-applications-to}}
\end{figure*}

The fight against the spread of the infection can benefit from pervasive
equipments that are currently adopted in daily life (i.e. smartphones
and wearable devices), namely, the rising framework of the Internet
of Things (IoT). This technology can be specialized in a fan of options
sharing same infrastructures, standards, and expertise \cite{Ding20}.
Different IoT protocols like Bluetooth Low Energy (BLE), Wireless
Fidelity (Wi-Fi), Global Positioning System (GPS), have already gained
popularity for providing solutions to pandemic-related challenges
in both domestic and hospital scenarios \cite{AdansDester20,Nasajpour20,Dong21}.
The Radiofrequency Identification (RFID), that is nowadays one of
the pillars of the Healthcare Internet of Things (H-IoT) \cite{Habibzadeh20},
is probably the most suitable technological approach to be scaled
up in order to support, by means of a single infrastructure, almost
all the logistic and monitoring issues related to the war against
a pandemic. RFID transponders \cite{Occhiuzzi13sensorless} can be
manufactured over large scale in any shape and material, including
plastics, paper, elastomers, as well as dissolvable organics compounds.
The tag can be embedded into things and can be attached onto the human
skin as well. Readers can be rugged and multifunction, integrable
into a smartphone and even into a smartwatch-like personal device.
Not least, RFID relies on standard protocols and is intrinsically
low-cost.

While the original purpose of RFID was the electronic labeling of
goods, it is nevertheless quickly moving to the monitoring of objects,
industrial processes, and people, by empowering low-cost devices with
sensing capabilities \cite{Occhiuzzi13sensorless}. Measurement of
temperature, pressure, humidity, just to list some, are now feasible
with commercial-off-the-shelf (COTS) tags \cite{farsens}. Advanced
applications, involving chemical parameters (pH, chlorides) and biophysical
signals (EMG, EKG, skin resistance), are under advanced investigation
in worldwide research labs \cite{Mazzaracchio20,Yap16}. Hence, there
is a creative wealth of technology all around that could be usefully
put in phase for application to pandemic emergencies.

The aim of this paper is, therefore, showing how the existing RFID
devices (technology introduced in Section II) and the most advanced
related research efforts could be synergically used to fight a COVID-19-like
pandemic. Three macro-topics (as schematized in Fig. \ref{fig:RFID-applications-to})
are considered: PPE supply, usage and management (Section III), Access
control, contact tracing and indoor ambient quality (Section VI),
and Predictive and Preventive Healthcare (Section V). For each topic,
the main issues and needs are resumed, and consequently the most appropriate
RFID devices and processes are identified. In some cases, the technology
has already been adopted in the proposed scenario. In other cases,
applications are in different fields but be employed in the fight
against pandemics. To improve readability and enable a direct access
to the required information and solutions, each of the three sections
can be considered as a standalone chapter. For each application, a
table lists the real-life problems and the possible solutions through
RFID technology. References are accordingly grouped to directly find
the source of the discussed topics.

The overall purpose of the paper is definitely to provide a survey
on what could be immediately applied to the pandemic-related problems,
as well as to draw a picture of the open challenges for the finalization
of current researches.

\section{UHF RFID Technology: Basic Principles and Latest Advancements}

\begin{figure}[t]
\begin{centering}
\includegraphics[width=8.5cm]{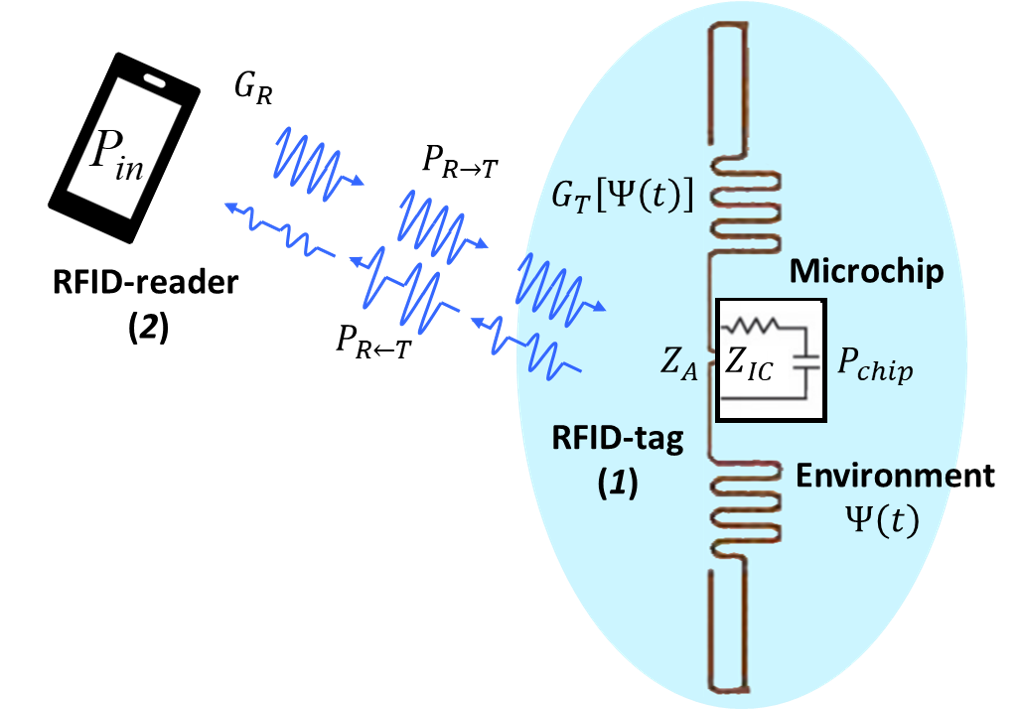}
\par\end{centering}
\caption{A sketch of the RFID system. A reader \emph{(2) }interrogates the
tag \emph{(1)}, the behavior of which is influenced by a changing
physical feature $\Psi$. Adapted from \cite{Occhiuzzi13sensorless}.\label{fig:RFID_scheme}}
\end{figure}

An RFID system (sketch in Fig. \ref{fig:RFID_scheme}) comprises two
components: \emph{(1) }the \emph{tag}, that is attached onto the object
(or even on the person) to be labeled, comprising an antenna and a
microchip transponder (Integrated Circuit, IC), and \emph{(2)} the
local querying system (or \emph{reader}), that energizes the tag from
remote and collects data reflected back from it throughout backscattering
modulation. The backscattering communication scheme applies for both
short range (namely HF/NFC systems working at $13.56$ MHz) and long
range (working in the UHF 860-960 MHz frequency band) platforms \cite{Dobkin:2007}.
In the former case the interaction with the reader is mostly one-to-one
(inductive coupling), whereas in the latter case a single reader can
simultaneously collect information from hundreds of tags, being the
radiating elements conventional antennas operating in far-field conditions.
Accordingly, the UHF band has the advantage to be tailored to control,
within a unitary framework, both single items and complex processes
\cite{Occhiuzzi19industry}, with extremely low power consumption
and maintenance cost (passive and battery-less systems). This paper
is therefore focused on systems working in the UHF frequency band,
which currently provide the largest amount of COTS devices for any
kind of material and are readable up to more than $15$ m \cite{Dobkin:2007}.

The tags can be passive, when they harvest energy from the interrogating
system, semi-active (BAP: Battery Assisted Passive), when a battery
is included to only feed the sensors and the internal memory, or fully
active. In the latter case, a local source directly feeds both a microcontroller
and the transmitting radio, such to sensibly increase the reading
distance and enable autonomous operations.

In passive and BAP technology (see again Fig. \ref{fig:RFID_scheme}),
at the beginning of the reader-to-tag communication protocol, the
reader first activates the tag by sending a continuous wave. It charges
an internal capacitor, providing the required energy to the tag to
perform actions. During the next steps of the communication, the tag
receives commands from the reader, and finally sends back the data
through backscattered modulation of the continuous wave provided by
the reader itself. In this case, the tag\textquoteright s IC acts
as a programmable switching device between a low impedance and a high
impedance state, thus modifying the reflectivity of the responding
tag itself, and hence the strength of the reflected power.

\begin{figure}[t]
\begin{centering}
\includegraphics[width=8cm]{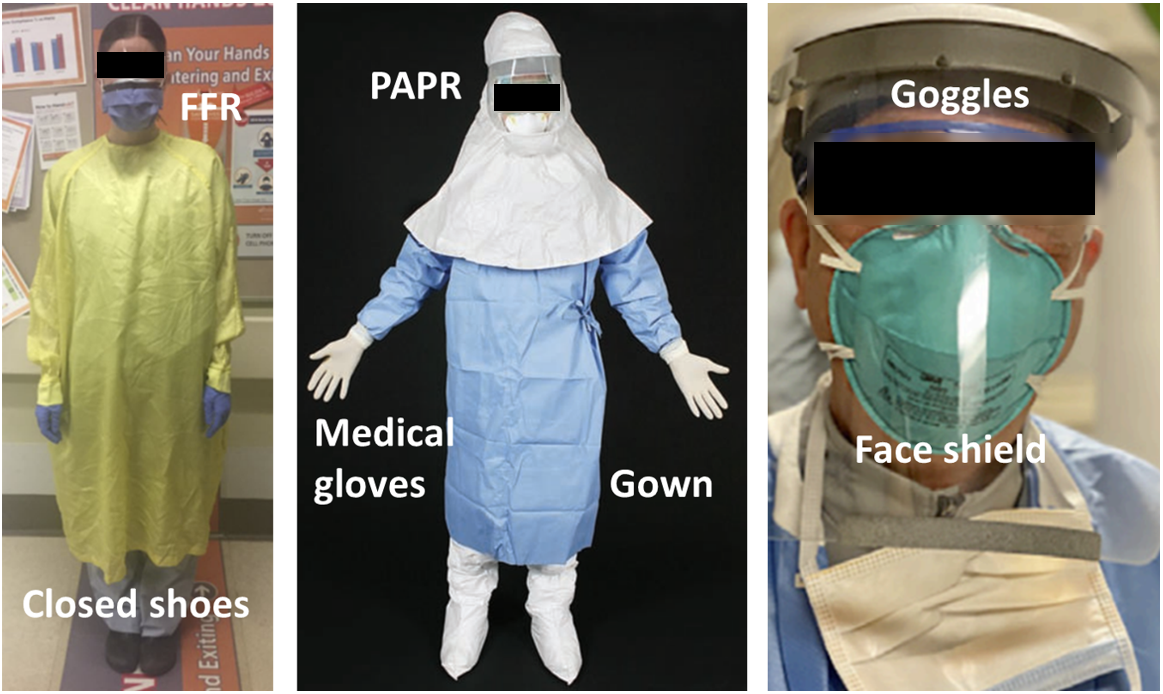}
\par\end{centering}
\begin{raggedright}
\ \ \ \ \ \ \ \ \ \ (a)~~\ \ \ \ \ \ \ \ \ \ \ \ \ \ \ \ (b)~~\ \ \ \ \ \ \ \ \ \ \ \ \ \ \ \ \ \ \ \ (c)\medskip{}
\par\end{raggedright}
\begin{centering}
\includegraphics[width=8cm]{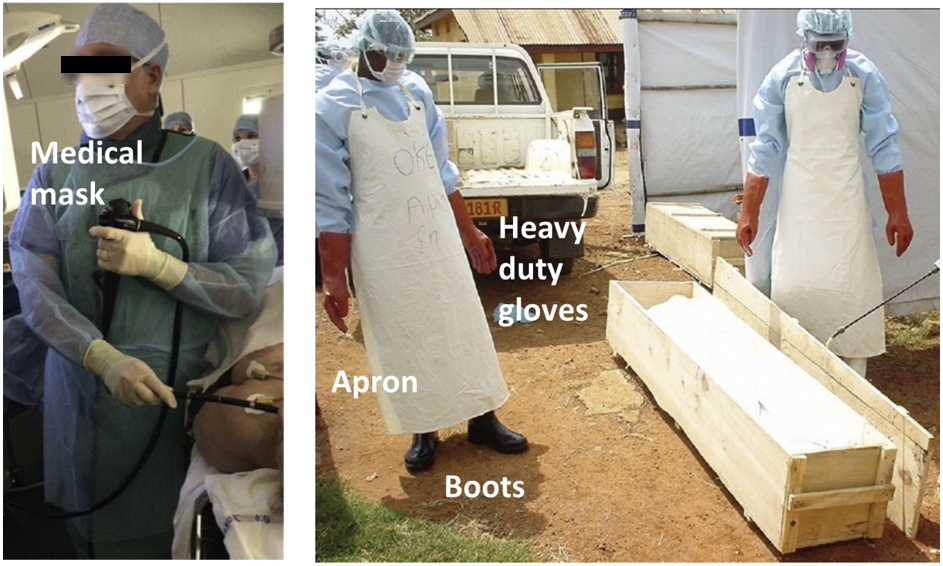}
\par\end{centering}
\begin{raggedright}
~~~~~~~~~~~~(d)~~~~~~~~~~~~~~~~~~~~~~\ \ \ \ ~~~~(e)
\par\end{raggedright}
\caption{Medical PPE used during the COVID-19 pandemic \cite{WHO2020}. (a)
FFR and closed shoes (adapted from \cite{Lockhart2020}). (b) PAPR,
medical gloves and gown (adapted from \cite{Zamora2006}). (c) Goggles
and face shield (adapted from \cite{Mostaghimi2020}). (d) Medical
mask (adapted from \cite{Alexandre2017}). (e) Heavy duty gloves,
apron and boots (adapted from \cite{RaabeVanessa2012}).\label{fig:All_PPE}}
\end{figure}

During a typical RFID communication, reader and tag share different
types of data:
\begin{itemize}
\item \emph{Electronic Product Code} (EPC) and \emph{User memory}, that
are the unique IDentification name assigned to the tag and up to 8k
byte of data; both can be writable many times and password protected;
\item \emph{Received Signal Strength Indicator }(RSSI), that is related
to the power backscattered by the tag toward the reader;
\item \emph{Turn-on power}, that is the minimum power the reader should
emit to wake-up the tag at a given distance;
\item \emph{Phase}, that is the phase of the signal backscattered by the
tag and related to the differential radar cross section $rcs$ among
the two modulating states.
\end{itemize}
The last three parameter are strictly related to the tag antenna operating
features. Any variation of the surrounding environment $\Psi$, able
to affect the antenna gain and input impedance, is transduced into
a variation of the signals transmitted and received by the reader,
and can be hence related to a sensing activity performed by the tag
itself. This analog approach, denoted as \emph{sensor-less sensing,}
can be applied to every passive tag if properly designed \cite{Occhiuzzi_design_2013}.
Extremely low-cost wireless sensors, namely disposable, can be obtained
at the price of a modest accuracy \cite{OcchiuzziAccuracy16}.

\begin{table*}[h]
\caption{Possible applications of tagged PPE during pandemics.\label{tab:PPE_RFID}}

\centering{}%
\begin{tabular}{>{\raggedright}p{4.5cm}>{\raggedright}p{4.5cm}>{\raggedright}p{5cm}>{\centering}p{2cm}}
\toprule 
\textbf{PPE problem} & \textbf{Possible RFID Application} & \textbf{Application examples} & \textbf{References}\tabularnewline
\midrule
\midrule 
\multirow{4}{4.5cm}{SUPPLY CHAIN MANAGEMENT} & PPE shortages & Localization and tracking inside hospitals to reduce lost and stolen
PPE & \centering{}\cite{QU2011}\cite{Mun2007}\cite{Hakim2006}\cite{Jarritt2014}\cite{Meiller2011}\cite{Ostbye2003}\cite{Shirehjini2012}\cite{Tsai2019}\tabularnewline
\cmidrule{2-4} \cmidrule{3-4} \cmidrule{4-4} 
 & Optimization of HC supply chains & Supply chains of hospitals and pharmacies & \centering{}\cite{Kumar2008}\cite{Coustasse2013}\cite{Smith2019}\cite{Borelli2013}\cite{Katsiri20162}\cite{Kalra2012}\tabularnewline
\cmidrule{2-4} \cmidrule{3-4} \cmidrule{4-4} 
 & \multirow{2}{4.5cm}{Anti-counterfeiting} & Supply chain integrity & \centering{}\cite{Cole2008}\cite{Inaba2008}\cite{SAFKHANI2020}\cite{Huang2010}\tabularnewline
\cmidrule{3-4} \cmidrule{4-4} 
 &  & Post-supply chain control through block-chain & \centering{}\cite{Raj2019}\cite{Toyoda2017}\tabularnewline
\midrule 
\multirow{2}{4.5cm}{APPROPRIATE USE OF PPE} & \multirow{2}{4.5cm}{Check if the HCW is wearing the proper PPE} & Gate-based check & \centering{}\cite{Bauk2018}\cite{KELM2013}\cite{Musu2014}\cite{Mahmad2016}\cite{Hung2019}\cite{Pradana2019}\cite{Manfred2012}\cite{Sole2013}\tabularnewline
\cmidrule{3-4} \cmidrule{4-4} 
 &  & A body-worn portable reader monitors the PPE worn by the worker & \centering{}\cite{BARROTORRES2012}\cite{Mandar2020}\cite{Cabreira2015}\cite{Smith2005}\tabularnewline
\midrule 
\multirow{3}{4.5cm}{PPE INTEGRITY} & \multirow{3}{4.5cm}{Correlate the PPE with sensed data about its conditions} & Maintenance of respirators optimized through their life cycle data & \centering{}\cite{Manfred2012}\tabularnewline
\cmidrule{3-4} \cmidrule{4-4} 
 &  & Pressure sensor to monitor the correct use of the PPE & \cite{DONG2018}\tabularnewline
\cmidrule{3-4} \cmidrule{4-4} 
 &  & Embedded humidity-sensing to monitor the effectiveness of the filtering
mask & \cite{Bianco21}\tabularnewline
\midrule 
\multirow{2}{4.5cm}{PPE DISPOSAL} & Monitor the disposal of the PPE (which is a medical waste) & Tracking the medical wastes to be disposed & \centering{}\cite{Liu2017}\cite{Liu2020}\cite{Katsiri2016}\cite{Nolz2014}\cite{Sun2019}\tabularnewline
\cmidrule{2-4} \cmidrule{3-4} \cmidrule{4-4} 
 & Improvements in the supply chain and in the reverse logistic of the
PPE to reduce the environmental footprint & Environmental footprint reduction by adopting the RFID in supply chains
and in the reverse logistic of medical wastes & \centering{}\cite{Bose2011}\cite{Liu20172}\cite{Gonzalez21}\tabularnewline
\midrule 
\multirow{2}{4.5cm}{HAND HYGIENE AND DISINFECTION} & \multirow{2}{4.5cm}{Ensure the compliance with the hand hygiene procedures} & Identify the HCW who needs to perform the hand hygiene & \cite{Bonanni2005}\cite{Do2009}\cite{Radhakrishna2015}\tabularnewline
\cmidrule{3-4} \cmidrule{4-4} 
 &  & Monitor the correction of the hand hygiene procedure & \cite{Decker2016}\cite{Johnson2012}\cite{Meydanci2013}\cite{Pineles2014}\cite{Pletersek2012}\tabularnewline
\bottomrule
\end{tabular}
\end{table*}

Nowadays, off-the-shelf RFID ICs with augmented sensing capabilities
are available \cite{onlineAXZON,onlineSL900,onlineEM4325,onlineAsygn}.
They include high-speed non-volatile memory (EEPROM), typically integrate
an embedded temperature sensor, and are provided with programmable
I/O ports for connecting general-purpose microcontrollers and sensors.
Sensing data are digitally encoded into the internal memory and transmitted
through standard backscattering communication. If compared with the
analog sensing, the collected physical information is definitively
more specific and robust \cite{Occhiuzzi13sensorless}. When provided
with battery, such ICs can also act as data-loggers, and they can
gather sensing data even without the need to receive energy from the
reader. These devices can be considered as a convergence point among
fully passive tags and the autonomous sensor nodes having local computational
capability with already assessed applications in Industry \cite{Occhiuzzi19industry}
and Healthcare \cite{Amendola14}.

\section{Personal Protective Equipment Management}

According to the WHO guidelines \cite{WHO2020}, HealthCare Workers
(HCWs) and general public must wear proper PPE depending on the activities
to be performed and on the occupied spaces. The set of appropriate
PPE to protect against pandemics (Fig. \ref{fig:All_PPE}) comprises:
\begin{itemize}
\item medical masks, filtration respirators (N95 or FFP2), cloth masks,
higher filtering masks, and Powered-Air Purifying Respirators (PAPR)
\cite{WHO2020_2,Holland2020};
\item gowns;
\item medical or heavy-duty gloves;
\item eye protections (e.g., goggles or face shields);
\item aprons;
\item boots or closed shoes.
\end{itemize}
In addition to these requirements, a proper hand hygiene is mandatory
\cite{WHO2020_2}.

The management of PPE requires:
\begin{enumerate}
\item an efficient supply chain and control of counterfeit products \cite{Ippolito2020,Pitts2020},
along with solutions to face their shortage, especially in the case
of misuse and overuse \cite{Cook2020};
\item to check the compliance of the HCWs with the PPE procedures and guidelines,
which moreover can be time-consuming or even confusing \cite{Houghton2020},
also considering the difficulty of working while wearing the PPE \cite{Hignett2020,Jiang2020};
\item to check the condition of PPE, as it overuse can cut down its effectiveness
\cite{Gao2016,Pacitto2019};
\item the care of the environmental impact produced by the wasted PPE, in
particular regarding their correct disposal \cite{Klemes2020}.
\end{enumerate}
Similar issues are related to the pharmaceutical chain, in particular
to vaccines and life-saving drugs \cite{WHOvaccines}, whose supply
and correct distribution is directly proportional to their effectiveness.

RFID systems that can be employed in the PPE management are resumed
in Table \ref{tab:PPE_RFID}.

\begin{figure*}[h]
\begin{centering}
\includegraphics[width=18cm]{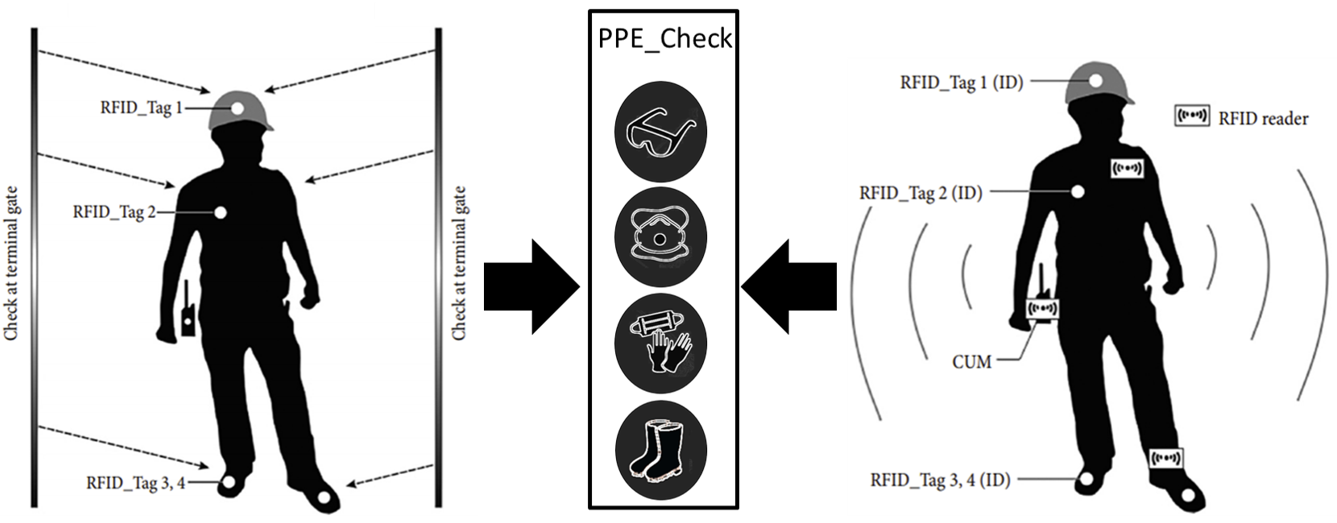}
\par\end{centering}
\begin{raggedright}
\ ~~~~~~~~~\ \ \ \ \ \ \ \ \ \ \ \ \ \ \ \ ~\ \ \ \ \ \ \ \ ~(a)\ \ \ \ \ \ ~~~~~~~~~~~~~~~~~~~~~~~~~\ \ \ \ \ \ \ \ \ \ \ \ \ \ \ \ \ \ \ \ \ \ \ \ \ \ \ \ \ \ \ \ \ \ \ \ \ \ \ \ \ ~(b)\medskip{}
\par\end{raggedright}
\begin{centering}
\includegraphics[width=17.9cm]{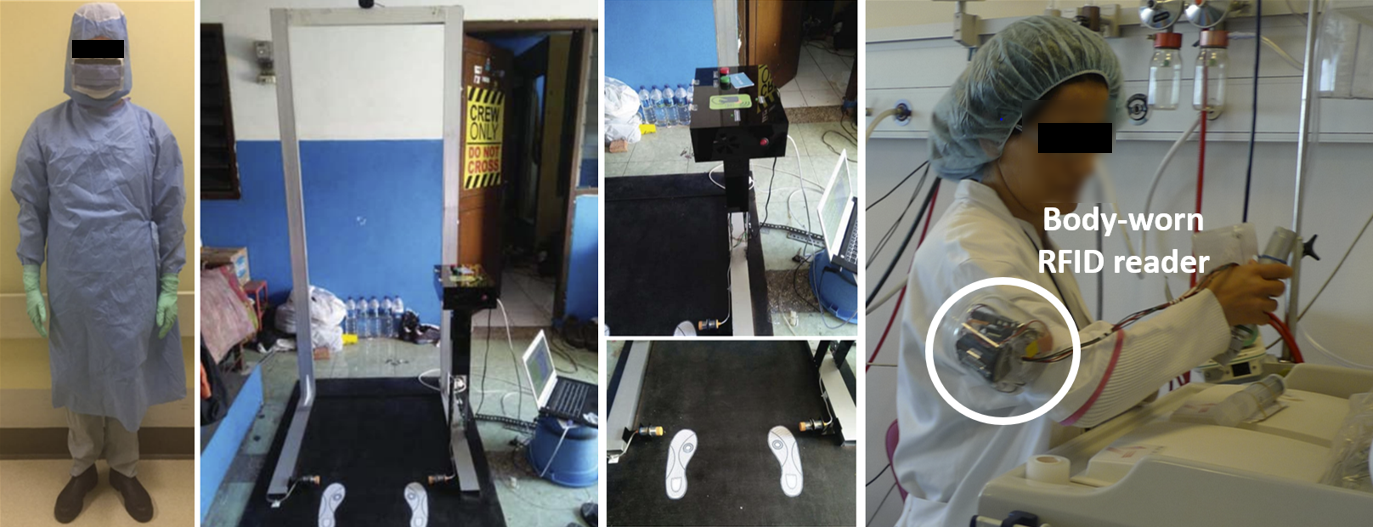}
\par\end{centering}
\begin{raggedright}
~~~~~~~~~~\ \ \ \ \ \ \ \ \ \ \ \ \ \ \ \ ~\ \ \ \ \ \ \ \ \ \ \ \ \ \ \ \ \ ~(c)~~~~~~~~~~~~~~~~~~~~~~~~~\ \ \ \ \ \ \ \ \ \ \ \ \ \ \ \ \ \ \ \ \ \ \ \ \ \ \ \ \ \ \ \ \ \ \ \ \ \ \ \ \ ~(d)
\par\end{raggedright}
\caption{System's architecture to monitor the PPE worn by the workers (adapted
from \cite{Bauk2018}): (a) terminals checking, eventually equipped
with cameras, and (b) body-worn readers checking through communication
with the Central Unit Microcontroller (CUM) embedded in the worker's
Personal Digital Assistant. Examples of implemented systems: (c) a
gate to monitor the compliance with the worn PPE (adapted from \cite{Lockhart2020},
\cite{Pradana2019}); (d) a portable RFID reader worn by a nurse (adapted
from \cite{Bardram2011}).\label{fig:PPE_Compliance}}
\end{figure*}

\subsection{Supply Chain Management}

RFID technology has been massively exploited for the supply chain
in healthcare \cite{Kumar2008} and pharmaceutical \cite{Kalra2012,SAFKHANI2020,Huang2010}
branches, with promising results in terms of cost reduction and patient
safety \cite{Kumar2008}.

\subsubsection{Shortages of PPE}

The shortages of PPE can be partly addressed by adopting RFID equipment-tracking
system within the hospital environment to increase their utilization
and reduce the number of lost and stolen items \cite{QU2011,Jarritt2014}.
RFID-based localization and tracking systems in healthcare scenarios
are resumed in \cite{Shirehjini2012,Denis2019,Gholamhosseini2019,Tsai2019}.
RFID can also be hybridized with either 2D-barcodes \cite{Mun2007}
or with infrared localization \cite{Ostbye2003}.

\subsubsection{Inventory of Vaccines}

Even though hospitals still widely employ barcodes to label equipments,
drugs and other medical items, RFID tags can be more effective thanks
to the possibility to store additional data on the tagged objects
\cite{Coustasse2013}, and, above all, for the potential upgrading
to sensing capabilities. In particular, RFID systems have already
been employed in the supply chain of the transfusion blood \cite{Borelli2013,Katsiri20162}
and in cold chain operations \cite{Vivaldi2020,Hu2020,Caizzone2011}.
Hence, they could be applied in the management of vaccines that need
to be kept at controlled temperature \cite{Mahase2020}.

\subsubsection{Anti-counterfeiting}

RFID tags could also be adopted for anti-counterfeit purposes \cite{Cole2008,Inaba2008},
in both pharmaceutical and PPE sectors \cite{SAFKHANI2020,Huang2010}.
The supply-chain integrity can be monitored by the synergistic action
of RFID and blockchain systems \cite{Raj2019}, especially when an
unambiguous ownership is missing \cite{Toyoda2017}, for example,
in most of the single-use medical PPE.

\subsection{Appropriate Use of PPE}

PPE is often cumbersome and difficult to use; therefore, HCWs may
not strictly follow the guidelines, consequently hindering the effectiveness
of the equipment. If the PPE is provided with an RFID tag, a reader
could check if the worker is wearing the proper pieces of equipment
\cite{BARROTORRES2012,KELM2013,Manfred2012,Sole2013}. A more robust
PPE check can be achieved by combining ID and sensing data. For instance,
tagged hardhats equipped with pressure sensors can ensure that the
worker is effectively wearing the equipment and not just carrying
it with him \cite{DONG2018}.

Overall, two kinds of architecture can be implemented (sketch in Fig.
\ref{fig:PPE_Compliance}) to check if the workers carry their PPE:\emph{
(i) }check-points, i.e. readers on gates and terminals (Fig. \ref{fig:PPE_Compliance}.a),
and \emph{(ii) }portable/wearable readers (Fig. \ref{fig:PPE_Compliance}.b).

\begin{figure}[t]
\begin{centering}
\includegraphics[width=8.5cm]{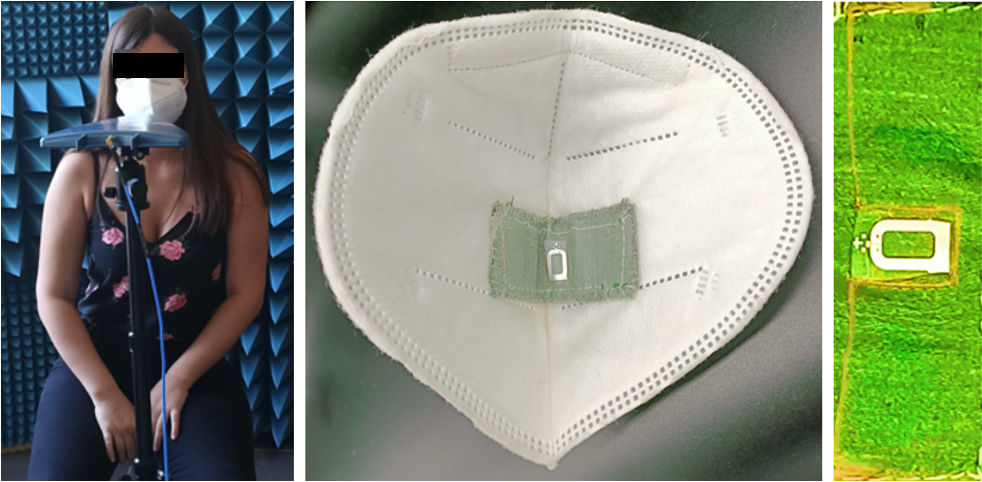}
\par\end{centering}
\begin{raggedright}
\ \ \ \ \ \ \ \ \ ~(a)~\ \ \ \ \ \ \ \ \ \ \ \ \ \ \ \ \ \ \ \ \ \ \ \ (b)\ \ \ \ \ \ \ \ \ \ ~~\ \ \ \ \ \ \ \ \ \ (c)\medskip{}
\par\end{raggedright}
\begin{centering}
\includegraphics[width=8.5cm]{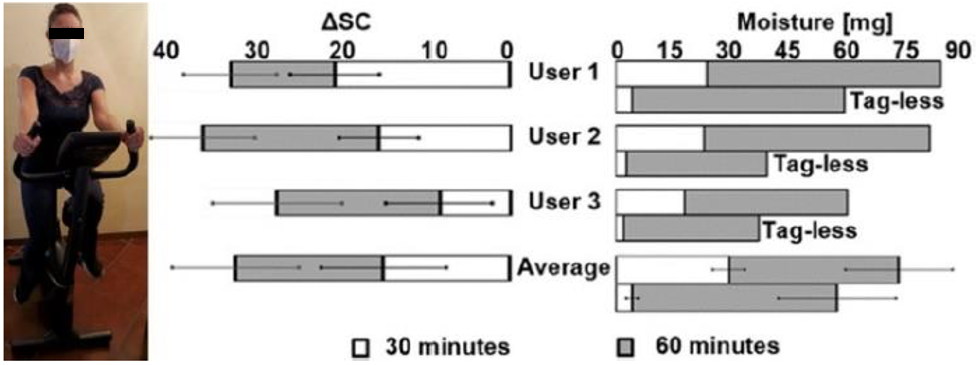}
\par\end{centering}
\begin{centering}
(d)
\par\end{centering}
\caption{Sensorized FFR. (a) Volunteer wearing a tagged facemask. (b) FFR-integrated
humidity tag. (c) Zoomed-in view of the textile humidity-sensing tag.
(d) Measured water moisture collected by the tagged and tag-less FFRs
after 30 and 60 minutes of physical exercise with a stationary bike.
Adapted from \cite{Bianco21}. \label{fig:mascherina}}
\end{figure}

\subsubsection{Gate-based Check}

Gate-based systems located in strategic points exploit RFID tags on
the PPE to either control the presence of the PPE itself when entering
dangerous areas \cite{Bauk2018,KELM2013,Musu2014,Mahmad2016,Manfred2012,Sole2013}
or\emph{ }to identify the worker while a camera checks for the PPE
compliance through images processing \cite{Hung2019,Pradana2019}.

\subsubsection{Body-worn Readers}

Since RFID gates cannot check if HCWs remove the PPE after the checkpoint,
portable readers can be adopted for continuous monitoring of the workers.
Readers could be either body-worn and connected to a Central Unit
Microcontroller (CUM) \cite{BARROTORRES2012,Cabreira2015}, or embedded
in the workers' Personal Digital Assistants (PDA) \cite{Mandar2020}.
Body-worn portable readers could also be used to recognize the activity
carried on by the HCW \cite{Bardram2011}, as the \emph{iBracelet}
in \cite{Smith2005}.

\begin{figure}[t]
\begin{centering}
\includegraphics[width=8.5cm]{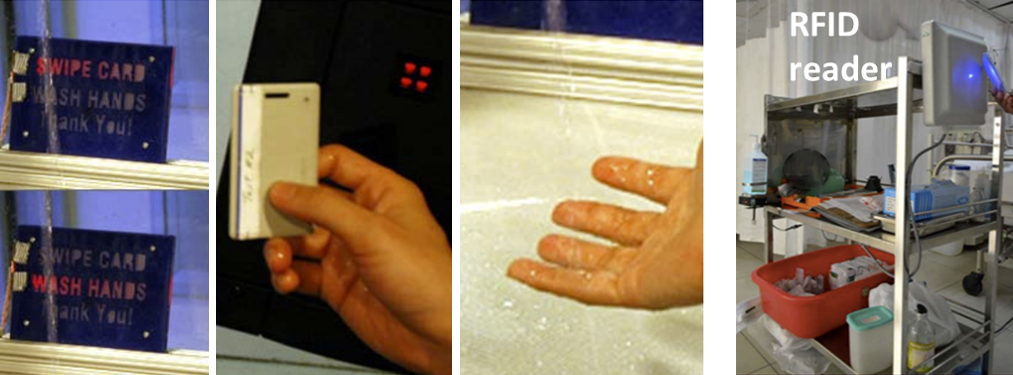}
\par\end{centering}
\ \ \ \ \ \ \ \ \ \ \ \ (a)\ \ \ \ \ \ \ \ \ \ \ \ \ \ \ \ \ \ \ \ \ \ \ \ \ \ \ \ \ \ \ ~~~(b)\medskip{}

\begin{centering}
\includegraphics[width=8.5cm]{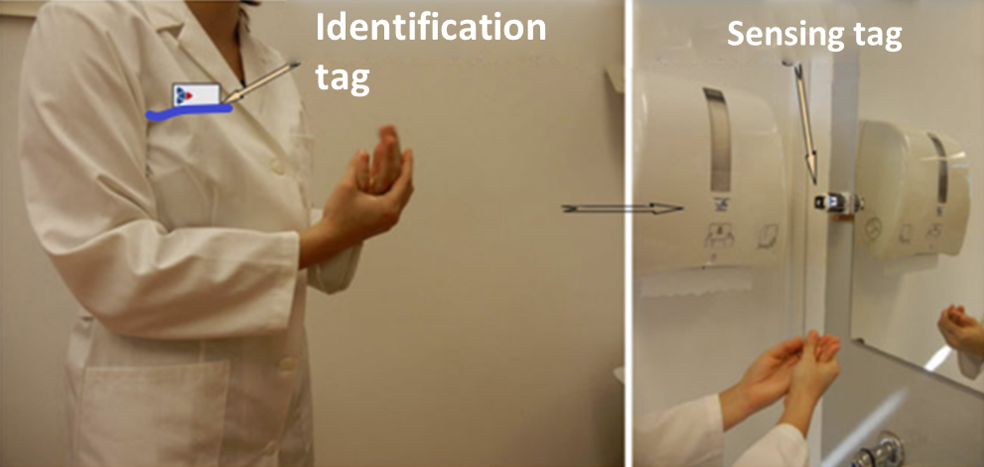}
\par\end{centering}
\begin{centering}
(c)
\par\end{centering}
\caption{RFID-based systems to monitor hand hygiene compliance. (a) A smart
sink identifying the user through RFID badge (adapted from \cite{Bonanni2005}).
(b) A dispenser of hand sanitizer exploiting RFID badges (adapted
from \cite{Radhakrishna2015}). (c) A system to monitor dispenser
activation, which couples the ID and the sensing information given
by two different tags (adapted from \cite{Pletersek2012}).\label{fig:mani}}
\end{figure}

\subsection{PPE Integrity}

The PPE health conditions must be in optimal conditions to guarantee
their effectiveness. As an example, since the filtering capabilities
of the facemasks are reduced by the breath humidity \cite{Gao2016,Pacitto2019},
the WHO advises replacing the masks as soon as they become damp \cite{WHO2020ter},
or at least after a specific use-time. An embedded humidity-sensing
tag is capable of discriminating normally used Filtering Facepiece
Respirators (FFRs) from abnormally wet ones (having more than $50$
mg of moisture on the internal surface; Fig. \ref{fig:mascherina})
\cite{Bianco21}. Moreover, RFID tags integrated inside the mask can
monitor the health status of the wearer \cite{Caccami2017}, as discussed
next in Section \ref{subsec:Respiratory-Function-and}.

\subsection{PPE Disposal}

A used PPE is an infectious medical waste that must be properly disposed
of, especially during pandemics \cite{WINDFELD2015}. The correct
disposal of wastes through RFID has been a well-studied topic in civil
and industrial sectors \cite{HANNAN2015} to reduce the environmental
impact \cite{Bose2011}. RFID could be effectively exploited also
for medical waste management \cite{Liu2017,Sun2019}. In such systems,
the tags are read thrice \cite{Katsiri2016,Liu2020}: when leaving
the hospitals, inside the trucks through truck-mounted readers, and
when arriving at the disposal site. The RFID can also be exploited
for the optimization of the route to collect the medical wastes \cite{Nolz2014},
with benefits also in recycling procedures and reverse logistics \cite{Liu20172}.

\begin{table*}[t]
\caption{RFID applications for patient management, contact control, access
control and the indoor ambient quality verification.\label{tab:Patient_Management}}

\centering{}%
\begin{tabular}{>{\raggedright}m{4.5cm}>{\raggedright}p{4.5cm}>{\raggedright}p{5cm}>{\centering}p{2cm}}
\toprule 
\textbf{Goal} & \textbf{RFID Application} & \textbf{Application examples} & \centering{}\textbf{References}\tabularnewline
\midrule
\midrule 
\multirow{4}{4.5cm}{ACCESS CONTROL} & Authorization verification & The individual entering a secured zone should be detected, and its
trustfulness should be automatically verified & \cite{Wu10}\cite{Pan12}\cite{Louati12}\cite{Shafin15}\cite{Bakht19}\cite{Amendola_Sept17_SCISSOR}\cite{Manzari12}\tabularnewline
\cmidrule{2-4} \cmidrule{3-4} \cmidrule{4-4} 
 & Occupants limitation, Individuals flow, and Resident time & Entrance/exit reports should be recorded to keep track of the check-in
or check-out status of each visitor & \cite{Shafin15}\cite{Bakht19}\cite{Buffi19}\cite{Amendola_Sept17_SCISSOR}\cite{Manzari12}\tabularnewline
\cmidrule{2-4} \cmidrule{3-4} \cmidrule{4-4} 
 & PPE-based authorization & Properly placed gates and terminals deny access if the entering individual
is not wearing the appropriate PPE & \cite{Manfred2012}\cite{Sole2013}\cite{KELM2013}\cite{Musu2014}\cite{Mahmad2016}\cite{Bauk2018}\cite{Hung2019}\cite{Pradana2019}\cite{Bianco21}\tabularnewline
\cmidrule{2-4} \cmidrule{3-4} \cmidrule{4-4} 
 & Temperature-based authorization & Properly placed gates and terminals deny access if an anomalous body
temperature is detected & \cite{Wen16}\cite{AmendolaBovesecchi16}\cite{iTek}\tabularnewline
\midrule
\multirow{5}{4.5cm}{PATIENT IDENTIFICATION AND MANAGEMENT} & \multirow{2}{4.5cm}{Labelling of the patient} & Assigning an ID to the patient & \cite{Chen2009}\cite{Chen20092}\cite{Chowdhury2007}\cite{Harry2014}\tabularnewline
\cmidrule{3-4} \cmidrule{4-4} 
 &  & Assign the patient's medical records to the tag & \cite{Perez2018}\tabularnewline
\cmidrule{2-4} \cmidrule{3-4} \cmidrule{4-4} 
 & \multirow{3}{4.5cm}{Tracking of the patient} & Univocally assign a reader to a room & \cite{Cao2014}\cite{Cheng2016}\cite{Chhetri2019}\cite{Iadanza2008}\cite{Magliulo2012}\cite{Mapa2015}\cite{Najera2011}\cite{Kim2008}\tabularnewline
\cmidrule{3-4} \cmidrule{4-4} 
 &  & Assign a tag to the location of the nearest reader & \cite{Newman-Casey2020}\cite{Kapoor21}\tabularnewline
\cmidrule{3-4} \cmidrule{4-4} 
 &  & Hybrid system exploiting RFID and Wi-Fi & \cite{Perez2017}\tabularnewline
\midrule
\multirow{4}{4.5cm}{CONTACT TRACING AND INTERPERSONAL DISTANCING} & \multirow{2}{4.5cm}{Contact tracing} & Readers in checkpoints & \cite{Bahri2007}\cite{Chang2011}\cite{Garg2020}\cite{North2013}\tabularnewline
\cmidrule{3-4} \cmidrule{4-4} 
 &  & Tag-to-tag communication & \cite{Cattuto2010}\cite{Stehle2011}\cite{Smieszek2016}\cite{Nikitin2012}\cite{Marrocco2012}\cite{Karimi2017}\cite{DeDonno2011}\cite{Piumwardane21}\tabularnewline
\cmidrule{2-4} \cmidrule{3-4} \cmidrule{4-4} 
 & \multirow{2}{4.5cm}{Social distancing} & Proximity detection & \cite{Borisenko2013}\cite{Bolic2015}\tabularnewline
\cmidrule{3-4} \cmidrule{4-4} 
 &  & Estimation of the tags' orientation to monitor face-to-face contacts & \cite{Isella2011}\cite{Saab2016}\cite{AlvarezNarciandi2018}\cite{Krigslund2011}\cite{Barbot2020}\cite{Akbar2019}\cite{Xiao2018}\tabularnewline
\midrule
\multirow{3}{4.5cm}{INDOOR MICROCLIMATE} & Room temperature & The monitoring of environmental temperature is carried out by deploying
multiple sensor units to constitute a USN & \cite{Cho05}\cite{Qi14}\cite{Occhiuzzi_Feb17_SCISSOR}

\cite{Amendola_Sept17_SCISSOR}\tabularnewline
\cmidrule{2-4} \cmidrule{3-4} \cmidrule{4-4} 
 & Room relative humidity (RH) & The monitoring of environmental humidity is carried out by deploying
multiple sensor units to constitute an USN & \cite{Oprea08}\cite{Siden09}\cite{Virtanen11}\cite{Manzari12_humidity}\cite{Amin13}\cite{Manzari13}\cite{Manzari14_J}\cite{Occhiuzzi_Feb17_SCISSOR}\cite{Amendola_Sept17_SCISSOR}\tabularnewline
\cmidrule{2-4} \cmidrule{3-4} \cmidrule{4-4} 
 & Air exchange (windows opening) & The occurrence of windows opening is enabled by properly distributed
light sensors and on-window transponders & \cite{Cho05}\cite{Colella16}\cite{Occhiuzzi_Feb17_SCISSOR}\cite{Amendola_Sept17_SCISSOR}\tabularnewline
\bottomrule
\end{tabular}
\end{table*}

\subsection{Hand Hygiene and Disinfection}

Hand hygiene is crucial to correctly wear and undress the PPE \cite{WHO2009,Dawson2014}.
Available RFID solutions are summarized in Fig. \ref{fig:mani}. RFID
can identify the HCW suggesting to perform the hand hygiene with a
given periodicity \cite{Bonanni2005,Do2009,Radhakrishna2015}. Then,
the HCW RFID badge can activate the hand sanitizer dispenser \cite{Decker2016,Johnson2012,Meydanci2013,Pineles2014}.
Alternatively, a sensor tag can monitor the dispenser activation by
measuring the propanol and ethanol concentration in the air \cite{Pletersek2012},
linked to the presence of the HCW ID in the same room.

\section{Access Control, Contact Tracing and Indoor Ambient Quality}

\begin{figure*}[h]
\begin{centering}
\includegraphics[scale=0.6]{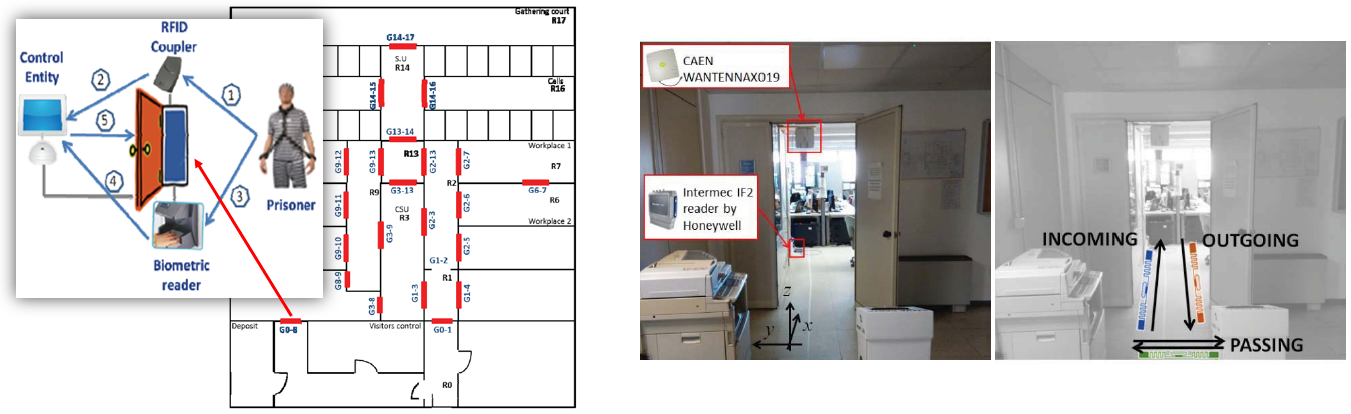}
\par\end{centering}
\begin{raggedright}
~~~~~~~~~~~~~~~~~~~~~~~~~~~~~~~~(a)~~~~~~~~~~~~~~~~~~~~~~~~~~~~~~~~~~~~~~~~~~~~~~~~~~~~~~~~~~~~~~~~~~~~~~(b)
\par\end{raggedright}
\begin{centering}
\includegraphics[scale=0.6]{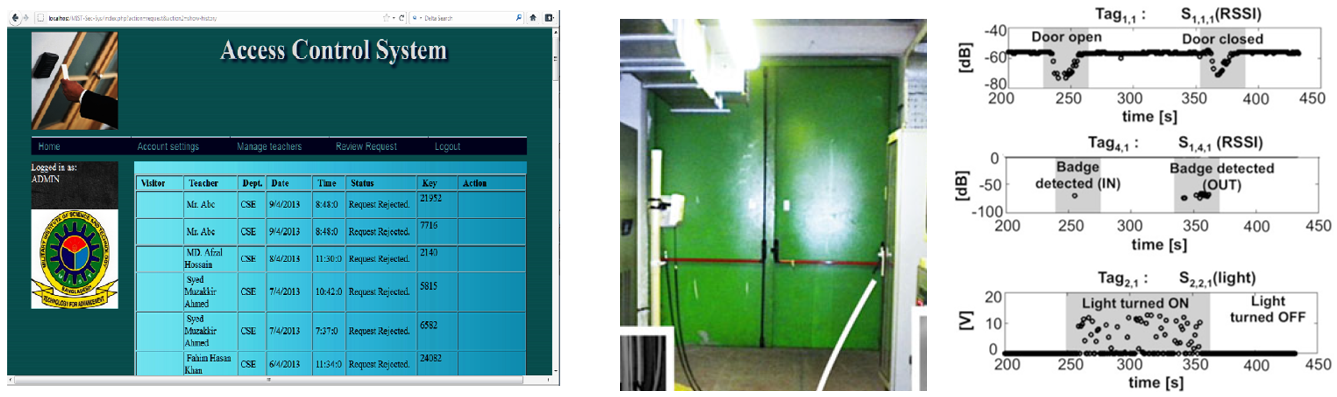}
\par\end{centering}
\begin{raggedright}
~~~~~~~~~~~~~~~~~~~~~~~~~~~~~~~~(c)~~~~~~~~~~~~~~~~~~~~~~~~~~~~~~~~~~~~~~~~~~~~~~~~~~~~~~~~~~~~~~~~~~~~~~(d)
\par\end{raggedright}
\caption{(a) Access control system testbed on a prison floor, with details
of biometric-based access control protocol (adapted from \cite{Louati12}).
(b) Gate access control setup at the door, able to discriminate the
flow of tagged individuals (adapted from \cite{Buffi19}). (c) RFID
card-based access control. Example of a summary of processed requests
(adapted from \cite{Shafin15}). (d) A subset of signals collected
by an RFID sensors network in case of authorized access to a restricted
area (adapted from \cite{Chapter_SCISSOR}).\label{fig:(a)-Access-control1}}
\end{figure*}

During pandemic emergencies, to allow a safe interaction among people,
especially in indoor spaces, three key strategies have been identified:
\emph{(i)} access control, \emph{(ii)} contact tracing, and \emph{(iii)}
social distancing. Mobile applications \cite{KRETZSCHMAR2020} and/or
wearable radios \cite{Smieszek2016} can be employed for interactions
monitoring. A specific case of access control and tracing is the management
of patients and staff in hospitals, points of care, and nursing homes,
especially in case of overwhelmed situations that could lead to delayed
cures \cite{Mareiniss2020,Maringe2020,Makoni2020}. In addition, since
infectious aerosols remain viable for up to 3 hours \cite{vanDoremalen20},
the simultaneous presence of people within a restricted environment
demands continuous air recycle. Moreover, if room temperature and
humidity are not appropriate, the risk of inhalation and infection
increases \cite{deMan20}. Hence, ventilation and air quality management
play a key role in preventing the spread of respiratory infections
indoors \cite{Spena20,Quraishi20}.

RFID-based access control, contact tracing and social distancing require
that people wear an RFID tag within their badge or on PPE. Furthermore,
since RFID tags empowered with sensing capabilities can enable the
possibility to get also the ``state'' of the tagged entity, the
monitoring of ambient quality becomes feasible too. Table \ref{tab:Patient_Management}
shows how RFID systems could be employed in the former scenarios,
as detailed next.

\begin{figure*}[h]
\begin{centering}
\includegraphics[width=18.2cm]{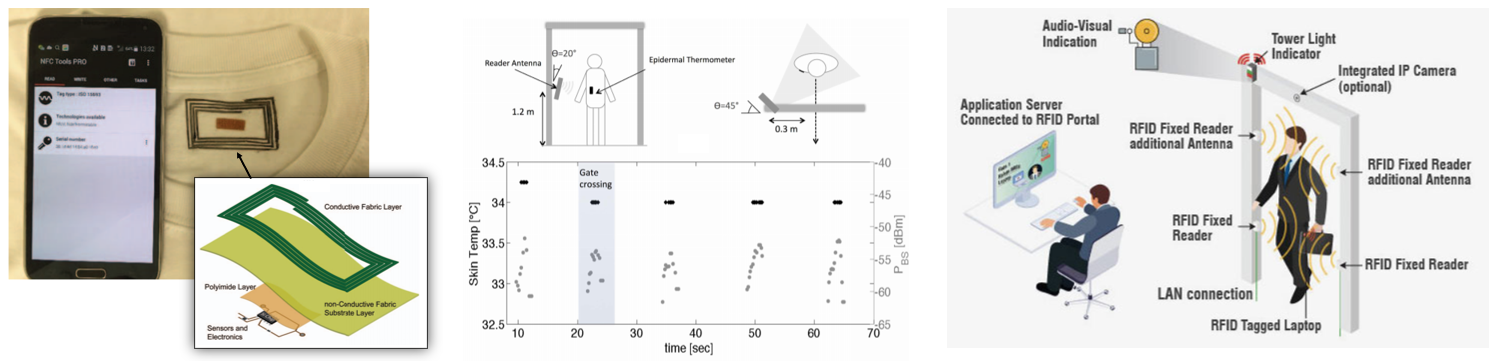}
\par\end{centering}
~~~~~~~~~~~~~~~~~~~~(a)~~~~~~~~~~~~~~~~~~~~~~~~~~~~~~~~~~~~~~~~~~~~(b)~~~~~~~~~~~~~~~~~~~~~~~~~~~~~~~~~~~~~~~~~~~~~~~(c)

\caption{(a) Wearable fabric-based RFID skin temperature monitoring patch,
glued on T-shirt (adapted from \cite{Wen16}). (b) RFID portal for
the automatic temperature screening of crossing people wearing epidermal
sensors (adapted from \cite{AmendolaBovesecchi16}). (c) Schematic
drawing of the automatic detection of temperature data of people during
the crossing of a surveillance gate (adapted from \cite{iTek}).\label{fig:access_control_2}}
\end{figure*}
\begin{figure}[ph]
\begin{centering}
\includegraphics[width=8.2cm]{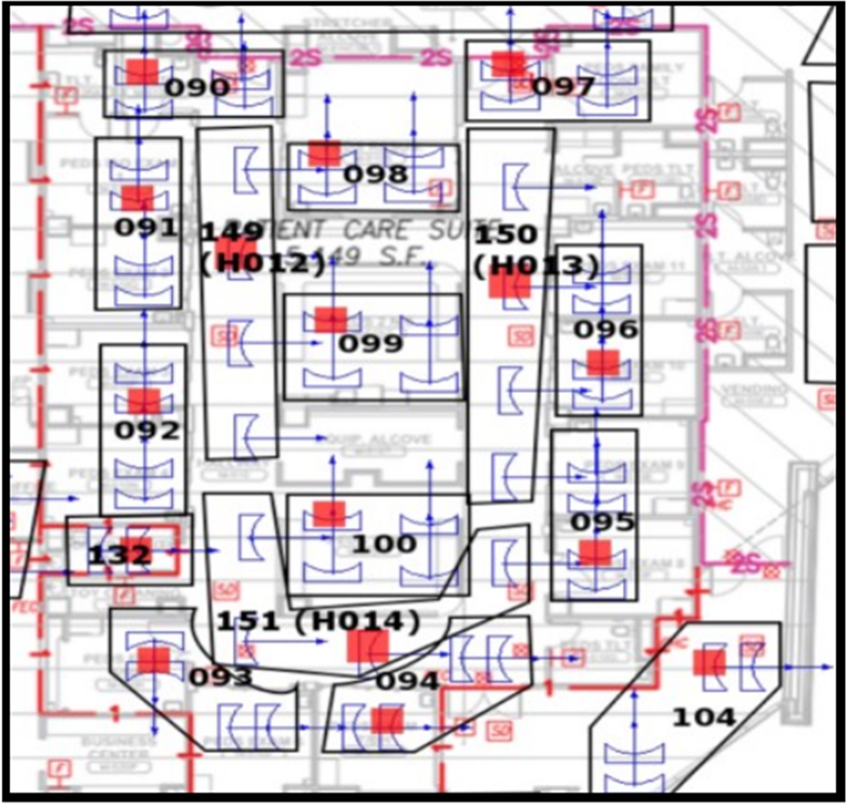}
\par\end{centering}
\begin{centering}
(a)
\par\end{centering}
\medskip{}

\begin{centering}
\includegraphics[width=8.4cm]{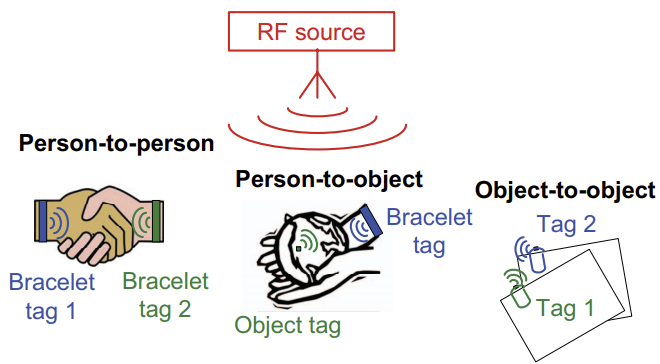}
\par\end{centering}
\begin{centering}
(b)
\par\end{centering}
\caption{(a) Map of RFID readers (blue sketches) installed in an emergency
department. The outlined boxes in the images are different areas (numbered
as reported near to the red squares), and multiple readers per area
are deployed (adapted from \cite{Kapoor21}). (b) Sketch of T2T systems,
wherein an illuminator allows the communication between tags (adapted
from \cite{Nikitin2012}). \label{fig:(a)-Concept-of}}
\end{figure}
\begin{figure}[t]
\begin{centering}
\includegraphics[width=8.7cm]{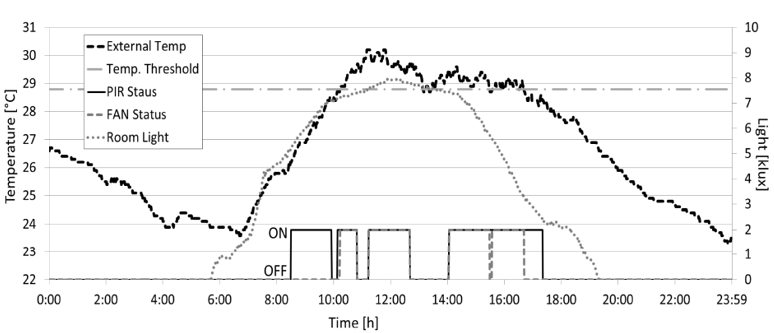}
\par\end{centering}
\caption{RFID-enabled ambient control on the basis of local data and context
data. Adapted from \cite{Colella16}.\label{fig:lux}}
\end{figure}

\subsection{Access Control}

Access control systems can leverage the RFID technology \cite{Bakht19}.
Regarding public areas, such as shops, restaurants, workplaces, classrooms,
and hospitals, access should be granted only if a set of criteria
is simultaneously satisfied: (\emph{i}) the person has the right privileges
to enter the restricted area (Fig. \ref{fig:(a)-Access-control1}.a);
(\emph{ii}) the number of people inside the venue does not exceed
the maximum capacity limit \cite{IndoorCapacity}; (\emph{iii}) the
individual correctly wears the required PPE, which at least includes
the protective facemask; (\emph{iv}) the body temperature is below
$37\ \textnormal{�C}$ \cite{CDCfever}.

\subsubsection{Detection of People Density}

RFID-based access control systems allow to count the number of people
inside space and also to generate log reports to keep track of check-in
and check-out status of each visitor \cite{Shafin15,Bakht19} (Fig.
\ref{fig:(a)-Access-control1}.c). RFID passive tags \cite{Manzari12}
can be integrated within the user badge and revealed by one or more
readers at the access gates for the automatic identification and the
storage of entrance/exit records. In particular, the monitoring of
people flow in closed areas can be achieved by applying phase-based
methods \cite{Buffi19} capable of discriminating incoming and outgoing
tags (Fig. \ref{fig:(a)-Access-control1}.b). By evaluating the time
spent inside the venue, the risk of exposure to infection can be estimated
accordingly.

In these applications, RFID tags can also be distributed in the environment
to be controlled so that the interaction with people will cause a
detectable change in their communication link to the reader. For instance,
the RFID device in \cite{Occhiuzzi_Feb17_SCISSOR} is used as a perimeter
sensor, placed on the floor and close to the access points, so that
variations in the returned RSSI can be exploited to detect/identify
the people passage. When deployed in a real-world application \cite{Marrocco_Oct16_SCISSOR,Amendola_Sept17_SCISSOR,Chapter_SCISSOR},
the RFID network is able to reveal unauthorized access in critical
areas (Fig. \ref{fig:(a)-Access-control1}.d), such as hospital wards
or vaccine storage warehouses.

\subsubsection{Preventing the Access of Non-Compliant People}

PPE-based access control systems should be implemented as a countermeasure
to deny access if the individual does not wear the required tagged
PPE (e.g., the facemask in Fig. \ref{fig:mascherina}, \cite{Bianco21}).
In addition, wearable tags with temperature sensing capabilities permit
to accomplish another easy countermeasure to deny access to potentially
ill individuals. The wearable tags can be integrated into clothes
(e.g. textile RFID tags \cite{Wen16,Merilampi16,Bjorninen15,Manzari12},
Fig. \ref{fig:access_control_2}.a) or even adhere directly on the
skin (e.g. epidermal RFID tags \cite{BookChapterAmendola}, see Fig.
\ref{fig:body_temp}, \ref{fig:axilla}, \ref{fig:fever}). Such a
temperature-based access control system enables the automatic detection
of the anomalous temperature of the individuals during the crossing
of a surveillance gate \cite{AmendolaBovesecchi16,iTek}, accordingly
triggering an alarm and/or preventing the doors from opening (Fig.
\ref{fig:access_control_2}.b,c).

\begin{figure*}[h]
\begin{centering}
\includegraphics[width=13.5cm]{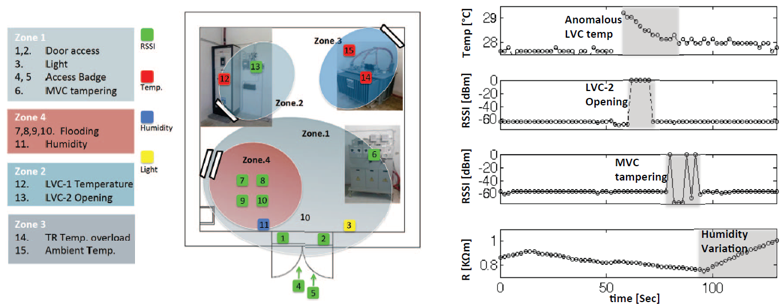}
\par\end{centering}
\begin{centering}
(a)
\par\end{centering}
\begin{centering}
\includegraphics[width=18.2cm]{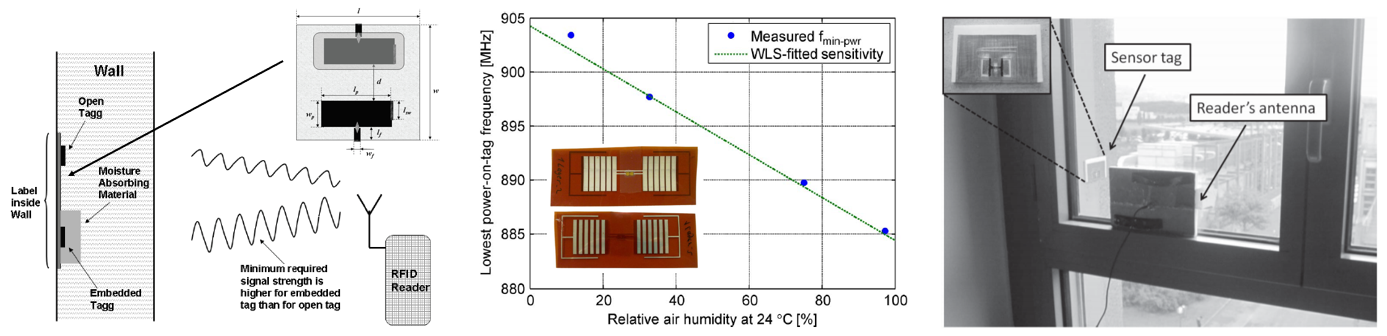}
\par\end{centering}
~~~~~~~~~~~~~~~~~~~~~~~~~~(b)~~~~~~~~~~~~~~~~~~~~~~~~~~~~~~~~~~~~~~~~~~~~(c)~~~~~~~~~~~~~~~~~~~~~~~~~~~~~~~~~~~~~~~~~~~~~(d)

\caption{(a) Architecture of an RFID sensors network for both ambient monitoring
and access control, with a subset of the measured electromagnetic
signals (adapted from \cite{Amendola_Sept17_SCISSOR}). (b) Moisture
sensing label incorporating two RFID tags to be integrated over walls
(adapted from \cite{Siden09}). (c) Calibration curve of the humidity
inkjet-printed sensor tag (adapted from \cite{Virtanen11}). (d) Measurement
setup for the overnight humidity exposure with a doped tag placed
outside an external wall and interrogated by an RFID reader from the
inside (adapted from \cite{Manzari12_humidity}).\label{fig:(a)-Moisture-sensing}}
\end{figure*}

\subsection{Patient Identification and Management}

\subsubsection{Identification}

Labeling of patients can be achieved by passive wristband tags, with
the aim of identifying them along clinical paths and optimizing the
clinical practices. RFID wristbands can be assigned before entering
the hospital, e.g. when accepted on the ambulance \cite{Chen2009}
or during the triage \cite{Chowdhury2007}. RFID-wristband can also
be applied to identify the pre-surgery patients as well as to manage
their clinical records \cite{Chen20092}. RFID systems, moreover,
permit to optimize the waiting times of the treatments \cite{Harry2014}
and also to keep track of medications within the Intensive Care Units
(ICUs). For instance, when an intravenous medication has to be administered
to the patient, the nurse can be required to validate the process
by reading together both the patient and the medication ID \cite{Perez2018}.

\subsubsection{Tracking}

Patients tracking can be achieved by linking one (or more) reader(s)
to one room so that the presence of the patients is retrieved from
the corresponding detected on-body tags \cite{Magliulo2012,Mapa2015}.
In this case, each reader is assigned to a specific area. Alternatively,
the assignment can be based on the maximum signal strength received
by the readers \cite{Cao2014} or by the tag \cite{Newman-Casey2020,Kim2008}.
In any case, the last reader illuminating the tag of the patient can
write its ID on the tag so that other readers can retrieve its last
known position \cite{Iadanza2008}. More accurate localization can
be obtained by active RFID tags that also exploit Wi-Fi signals for
the localization of the tagged individual \cite{Perez2017}.

\subsection{Contact tracing and Interpersonal Distancing}

RFID-based contact tracing is based on \emph{(i}) readers placed in
checkpoints (Fig. \ref{fig:(a)-Concept-of}.a), or \emph{(ii}) tag-to-tag
(T2T) interaction (Fig. \ref{fig:(a)-Concept-of}.b).

\subsubsection{RFID-based Check-Points}

This first architecture is similar to the standard people tracking
strategy, wherein people are tagged, and properly placed readers monitor
specific areas. RFID systems were successfully deployed in Taiwan
and Singapore \cite{Bahri2007,Wang06} during the epidemic outbreak
of the Severe Acute Respiratory Syndrome (SARS). In both cases, the
systems were based on active UHF RFID tags and differed only for the
people that were tagged: in Singapore everyone entering the hospital
was tagged; instead, in Taiwan, only the medical staff. It is also
possible to place a reader on each patient's bed to identify approaching
nurses and physicians. This system can successfully track almost every
interaction up to a range of $1.75$ m and can prevent nosocomial
infections \cite{Chang2011,VanPraet2020}.

\subsubsection{Face-to-Face Tracing}

The current bottleneck of massive implementation of RFID-based contact
tracing is the high cost of readers. This critical issue could be
overcome by resorting to the architecture of Tag-to-Tag (T2T) communication
\cite{Nikitin2012}, which uses a much simpler reader, namely a continuous
wave generator. T2T communication is based on \emph{reader tags} (also
known as \emph{talkers }\cite{Marrocco2012}), capable of interrogating
\emph{listener tags }by backscattering the field generated by an RF
source (the \emph{illuminator} \cite{Nikitin2012}). When tags of
different items get in close proximity, they start interacting and
sharing information, hence recording the meeting event. Although the
T2T maximum achievable communication distance can be as low as $25$
mm \cite{Nikitin2012,Marrocco2012}, a demodulator capable of extending
the T2T communication range up to $1.5$ m has been developed \cite{Karimi2017}.
A different kind of T2T communication involves receiving-only RFID
listeners, called \emph{augmented RFID receiver }(ARR), creating an
\emph{augmented RFID system }that monitors the contacts of tagged
people near the listener \cite{DeDonno2011}, up to a distance of
$3$ m \cite{Borisenko2013}. A type of ARR named \emph{Sense-a-Tag}
can be embedded into a wristband to sense the interactions of the
user with tagged objects \cite{Bolic2015} and possibly detect face-to-face
contacts that are known to require extra-distancing between people
to prevent airborne infections \cite{ECDC2020}.

\begin{table*}[h]
\caption{Wearable RFID Sensors for the monitoring of positive or suspected
infected individuals.\label{tab:Wearable-technology}}

\centering{}%
\begin{tabular*}{18cm}{@{\extracolsep{\fill}}>{\raggedright}p{2.5cm}>{\raggedright}p{2.2cm}>{\raggedright}p{8.35cm}>{\centering}p{2.3cm}>{\centering}p{1cm}}
\toprule 
\textbf{Clinical Evaluation of COVID-19} & \textbf{Measured Parameters} & \textbf{Wearable RFID Sensors Potentials} & \textbf{References} & \textbf{Phases}\tabularnewline
\midrule
\midrule 
\multirow{2}{2.5cm}{CLINICAL SYMPTOMS MONITORING} & Body Temperature & \textbullet{} Comfortable on-skin plaster-like devices

\textbullet{} Continuous or periodic measurements of body temperature

\textbullet{} Fever detection & \cite{iTek}\cite{Camera20}\cite{Occhiuzzi20}\cite{Miozzi20thin}\cite{Camera19}\cite{Miozzi19}\cite{Miozzi18}\cite{Miozzi18trial}\cite{Miozzi17}\cite{AmendolaBovesecchi16}\cite{Adame16}\cite{Milici14}\cite{Wang06} & I, II\tabularnewline
\cmidrule{2-5} \cmidrule{3-5} \cmidrule{4-5} \cmidrule{5-5} 
 & Cough & \textbullet{} Fabric-integrated devices

\textbullet{} Coughing fit detection

\textbullet{} Indirect breath monitoring (RSSI variations) & \cite{Manzari12} & I, II\tabularnewline
\midrule 
\multirow{2}{2.5cm}{RESPIRATORY ASSESSMENT} & Respiratory Rate,

Respiratory Depth,

Respiratory Cycle,

Breath Effort & \textbullet{} Comfortable facemask or on-skin plaster-like devices,
or fabric-integrated devices

\textbullet{} Abnormal respiration patterns detection (e.g. apnea,
tachypnea, ...)

\textbullet{} Direct (breath temperature or humidity) or indirect
(phase or RSSI variations) breath measurements & \cite{Chang21}\cite{Miozzi21giorgia}\cite{Oliveira20}\cite{Wang19}\cite{Hussain19}\cite{Caccami18breathanomalies}\cite{Occhiuzzi18}\cite{Caccami18}\cite{Araujo18}\cite{Hui18}\cite{Sharma18}\cite{Caccami17}\cite{Hou17}\cite{Hu17} & I, II\tabularnewline
\cmidrule{2-5} \cmidrule{3-5} \cmidrule{4-5} \cmidrule{5-5} 
 & Blood Oxygen Saturation & \textbullet{} Continuous pulse oximetry monitoring

\textbullet{} Very low power consumption & \cite{Wang17oximetry} & II\tabularnewline
\midrule 
\multirow{2}{2.5cm}{CARDIOVASCULAR STATUS} & ECG,

Hearth Rate & \textbullet{} Comfortable on-skin or fabric-integrated devices

\textbullet{} Continuous or periodic hearth rate and multichannel
ECG monitoring

\textbullet{} Very low power consumption & \cite{Horne20}\cite{Wang18}\cite{Hui18}\cite{Agezo16}\cite{Adame16}\cite{Vora15}\cite{Wang13}\cite{Besnoff13} & II\tabularnewline
\cmidrule{2-5} \cmidrule{3-5} \cmidrule{4-5} \cmidrule{5-5} 
 & Cuffless Blood Pressure & \textbullet{} Fabric-integrated devices

\textbullet{} Continuous blood pressure monitoring in static condition
(in care facilities) & \cite{Hui18} & II\tabularnewline
\midrule
\multirow{2}{2.5cm}{PSYCHO-PHYSICAL STRESS MONITORING} & Sweat & \textbullet{} Comfortable on-skin plaster-like devices or fabric-integrated
devices

\textbullet{} Continuous or periodic sweat pH, biomarkers and compound
measurements

\textbullet{} Psycho-physical stress detection

\textbullet{} Direct (sweat chemicals) or indirect (RSSI variations)
sweat measurements & \cite{Mazzaracchio20}\cite{Miozzi19}\cite{Nappi19}\cite{Merilampi16} & I\tabularnewline
\cmidrule{2-5} \cmidrule{3-5} \cmidrule{4-5} \cmidrule{5-5} 
 & Sleep Disorders & \textbullet{} Fabric-integrated and ambient tags

\textbullet{} Continuous overnight sleep quality evaluation

\textbullet{} Overnight body movements detection

\textbullet{} Possible integration with other RFID sensors (e.g. body
temperature, sweat compound, ...) & \cite{Hu17}\cite{Hussain19}\cite{Sharma18}\cite{AmendolaBovesecchi16}\cite{Occhiuzzi14}\cite{Occhiuzzi14night}\cite{Occhiuzzi10} & I, II\tabularnewline
\bottomrule
\end{tabular*}
\end{table*}

\subsection{Indoor Microclimate}

The American Society of Heating, Refrigerating and Air-Conditioning
Engineers (ASHRAE) suggests doubling the outside air changes per hour
with respect to the recommended ventilation rate before the pandemic
($14\,\textnormal{\ensuremath{\textnormal{m}^{3}}/hr}$) \cite{online_ventilation}.
Moreover, indoor temperature and relative humidity (RH) should be
maintained within determined ranges (e.g., $20$-$24\ \textnormal{�C}$
and $40$-$60\,\%$ respectively) to mitigate the virulence of infectious
agents \cite{Quraishi20,Spena20}.

\subsubsection{Windows Opening}

RFID tags embedding light sensors can be used to monitor the windows
opening to foster the exchange of air \cite{Cho05,Colella16,Occhiuzzi_Feb17_SCISSOR}.
For instance, by sensing the light level, a fan located in an office
room can be activated only when certain conditions are verified \cite{Colella16}
(Fig. \ref{fig:lux}). More elementary strategies can involve passive
tags disposed on the windows, whose opening/closure is retrieved by
processing the RSSI from the tags \cite{Chapter_SCISSOR}.

\subsubsection{Indoor Temperature and Humidity}

The microclimate of a room, namely the temperature and humidity (Fig.
\ref{fig:(a)-Moisture-sensing}.a), can be monitored by sensor-oriented
tags, thus providing a further indication about critical environmental
conditions that require air recycling. Environmental temperature can
be monitored by means of battery-less UHF RFID transponders provided
with a temperature sensor \cite{Cho05,Qi14}. Humidity sensors can
be instead designed providing the RFID tag with a material having
a high ability to absorb moisture and changing its dielectric properties
accordingly \cite{Oprea08,Siden09,Virtanen11,Manzari12_humidity,Amin13,Manzari13,Manzari14_J}
(Fig. \ref{fig:(a)-Moisture-sensing}.b,c,d). The same reader infrastructure
deployed for the access control can be used for the interrogation
of the above environmental sensors.

\section{Predictive and Preventive Healthcare\label{sec:Preventive-Healthcare}}

The health monitoring of suspected or confirmed infected people can
be referred to three different and often consecutive phases: (I) Preventive
early diagnosis, (II)\emph{ }Domestic disease monitoring, (III) Hospitalization.

In the former phase, preventive testing programs, both viral (swabs)
and physical, should be performed. Examples are temperature monitoring
\cite{CDCfever}, cardiovascular and respiratory screening (i.e. cardiac
rate, cough and altered respiration \cite{CDC}, and oxygen saturation
\cite{Petrilli20}), especially in case of individuals belonging to
controlled and restricted communities, such as hospitals, nursing
homes, assisted living facilities, factories and workplaces in general,
schools and prisons \cite{Hasell20}.

In the second and third phases, the health status of isolated individuals
should be continuously monitored to intervene in a timely manner in
case of rapid worsening. For instance, the domestic monitoring of
symptoms of COVID-19 positive patients for signs of deterioration
has demonstrated to effectively decrease the mortality rates \cite{Sun20}
and the hospital loads in China.

\begin{figure}[t]
\begin{centering}
\includegraphics[width=8.8cm]{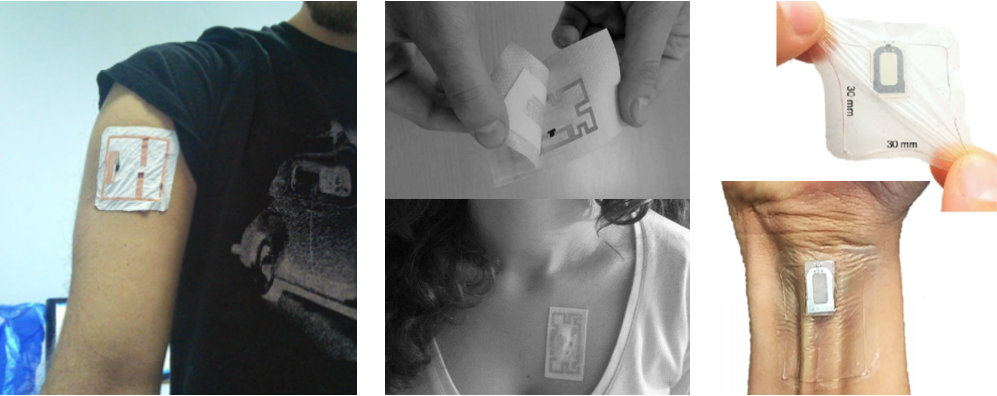}
\par\end{centering}
\begin{raggedright}
~~~~~~~~~~~(a)~~~~~~~~~~~~~~~~~~~~~
(b) ~~~~~~~~~~~~~~~~~~~(c)
\par\end{raggedright}
\begin{centering}
\includegraphics[width=8.8cm]{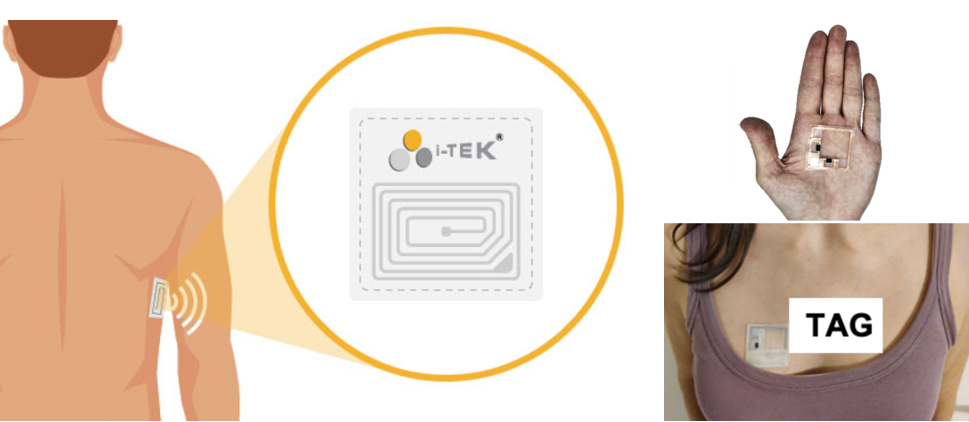}
\par\end{centering}
\begin{raggedright}
~~~~~~~~~~~~~~~~~~~~~~~~(d)~~~~~~~~~~~~~~~~~~~~~~~~~~~~~~~
(e)
\par\end{raggedright}
\caption{On-skin temperature RFID sensors. (a) Worn prototype of the epidermal
RFID temperature sensor (adapted from \cite{Milici14}). (b) Flexible
epidermal RFID thermometer (adapted from \cite{Miozzi18trial}, \cite{AmendolaBovesecchi16}).
(c) Flexible and conformable RFID thermometer (adapted from \cite{Miozzi20thin}).
(d) i-TEK's body temperature monitoring system (adapted from \cite{iTek}).
(e) Dual-chip flexible epidermal RFID temperature device (adapted
from \cite{Occhiuzzi20}). \label{fig:body_temp}}
\end{figure}

Wearable and epidermal RFID sensors \cite{BookChapterAmendola} exhibit
a great potential to enable a comfortable, easy-to-use and wireless
monitoring of a significant set of physiological parameters, such
as temperature, breath anomalies, electrophysiology, and biomarkers
in sweat, through small devices to be applied directly over the skin
\cite{Caccami18epidermal}, \cite{Panunzio21covid}. Table \ref{tab:Wearable-technology}
summarizes the possible applications of RFID platforms versus some
symptoms of the infectious disease \cite{Guan20}.

\subsection{Body Temperature}

During the fight against the SARS emergency in 2006, UHF active tags
with an embedded thermometer, employed for the Location-based Medicare
Service (LBMS) project at the Taipei Medical University Hospital \cite{Wang06},
demonstrated that the automatic and real-time temperature taking reduces
the risk of staff infections by limiting the contact with patients.
This system, moreover, improved patients safety by detecting fever
events in a timely manner.

\begin{figure}[t]
\begin{centering}
\includegraphics[width=8.8cm]{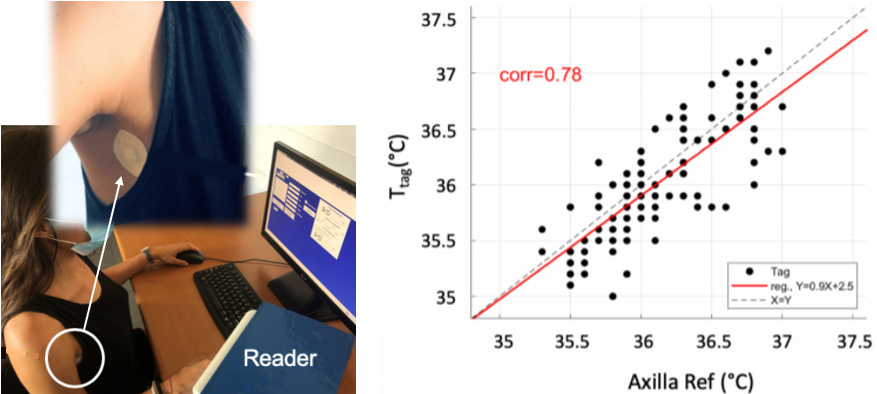}
\par\end{centering}
\caption{Plaster-like under-arm RFID thermometer. Correlation with an axilla
electronic thermometer. Adapted from \cite{Camera20}. \textcolor{blue}{\label{fig:axilla}}}
\end{figure}

RFID epidermal thermometers are suitable to be directly attached to
the human skin by means of bio-compatible transpiring membranes \cite{Milici14}
(Fig. \ref{fig:body_temp}.a). Over the years, they have become increasingly
more flexible, conformable \cite{AmendolaBovesecchi16} (Fig. \ref{fig:body_temp}.b),
\cite{Miozzi20thin} (Fig. \ref{fig:body_temp}.c), and small \cite{Camera20}
(Fig. \ref{fig:axilla}), also paying attention to their reusability
to reduce fabrication costs and pollution \cite{Miozzi17}. New pioneering
studies \cite{Occhiuzzi20} look in the direction of multichip RFID
temperature sensors to strengthen the correlation between the skin
and core body temperature (Fig. \ref{fig:body_temp}.e).

\begin{figure}[t]
\begin{centering}
\includegraphics[width=8.8cm]{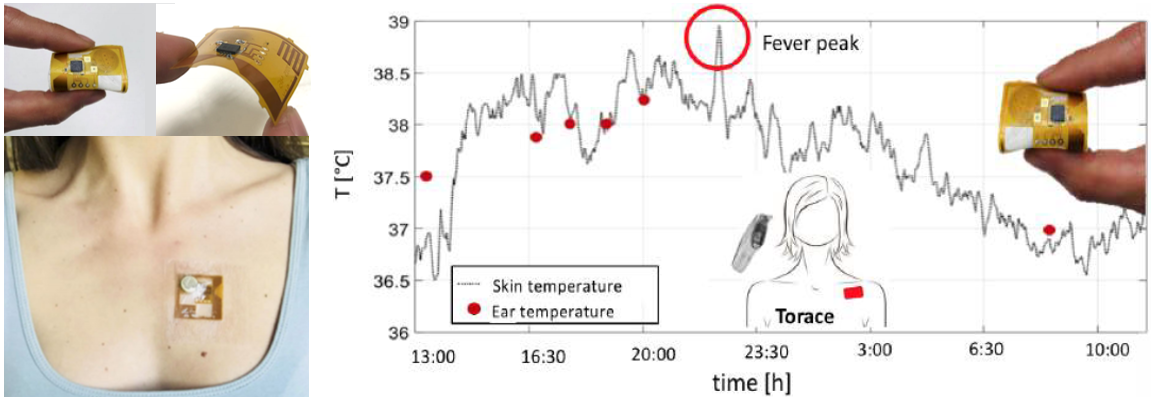}
\par\end{centering}
\caption{Small-size flexible RFID BAP epidermal sensor board and continuous
temperature monitoring against sporadic tympanic sampling.\textcolor{blue}{{}
}Adapted from \cite{Miozzi19,Miozzi18}. \textcolor{blue}{\label{fig:fever}}}
\end{figure}

Passive RFID epidermal thermometers can be read up to $1.5$-$2\,m$
\cite{Panunzio21SS}, hence being compliant with an automatic and
on-the-fly reading through gates and check-points to enable the unsupervised
detection of anomalous temperature peaks \cite{iTek} (Fig. \ref{fig:body_temp}.d).
Moreover, envisaged applications are also in clinical and domestic
settings by means of fixed remote antennas (e.g. continuous overnight
temperature monitoring). In this sense, these thermal sensors have
already been experimented on a hospital ward \cite{Miozzi18trial}
and have been deemed robust and reliable. If placed in the armpit
region (Fig. \ref{fig:axilla}), they are well correlated with a standard
axilla electronic thermometer (Pearson's coefficient $p=0.78$) with
a difference of less than $0.6\ ^{\textnormal{o}}\textnormal{C}$
in the $95\%$ of measurements \cite{Camera20}.

Continuous sampling of the skin temperature can also be achieved by
exploiting BAP RFID tags \cite{Miozzi19} (Fig. \ref{fig:fever})
working as data-loggers. Their validity was demonstrated in \cite{Miozzi18}
for the measurement of the skin temperature in realistic conditions
and in \cite{Camera19} for the evaluation of thermal stress of firefighters
during training.

\subsection{Respiratory Function and Cough \label{subsec:Respiratory-Function-and}}

\subsubsection{Cough}

\begin{figure}[t]
\begin{centering}
\includegraphics[width=8.8cm]{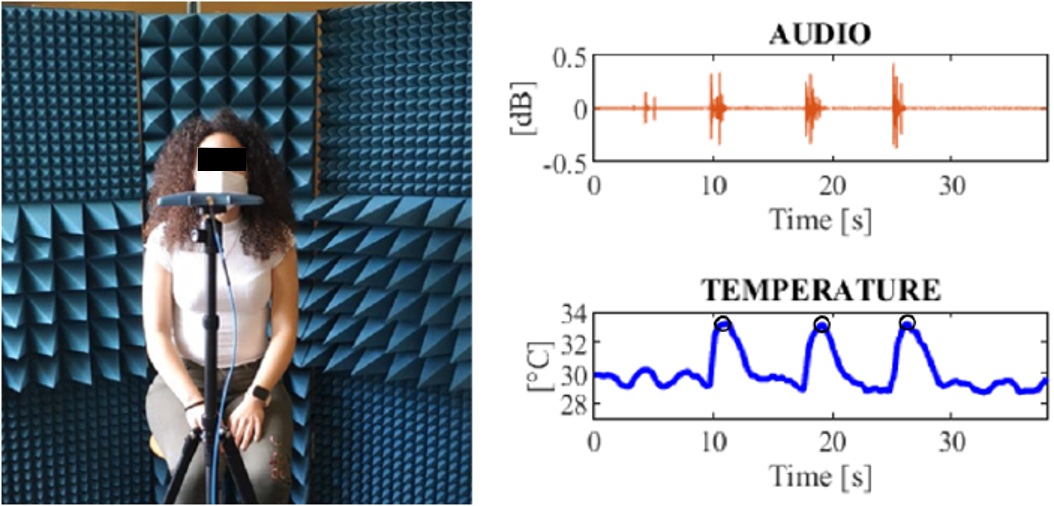}
\par\end{centering}
\caption{RFID-enabled cough evaluation system through detection of temperature
spikes. Adapted from \cite{Bianco21RFIDResearch}.\label{fig:tosse}}
\end{figure}

The presence and the dynamics of coughing can be monitored by wearable
RFID devices sewed into clothes on the chest region (Fig. \ref{fig:breath_sens}.c),
exploiting the backscattered power in combination with motion inertial
sensors \cite{Manzari12}. Another simple mean to count cough events
is based on sensorized facemasks that enable the detection of cough's
periodicity and duration by detecting the temperature spikes produced
by cough shots \cite{Bianco21RFIDResearch} (Fig. \ref{fig:tosse}).

\begin{figure}[t]
\begin{centering}
\includegraphics[scale=0.48]{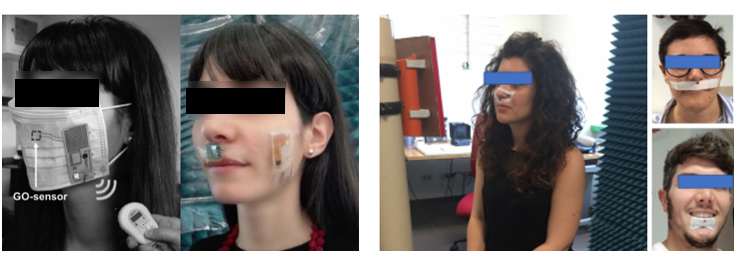}
\par\end{centering}
(a) ~~~~~~~~~~~~~~~~~~~~~~~~~~~~~~~(b)
\begin{centering}
\includegraphics[width=8.8cm]{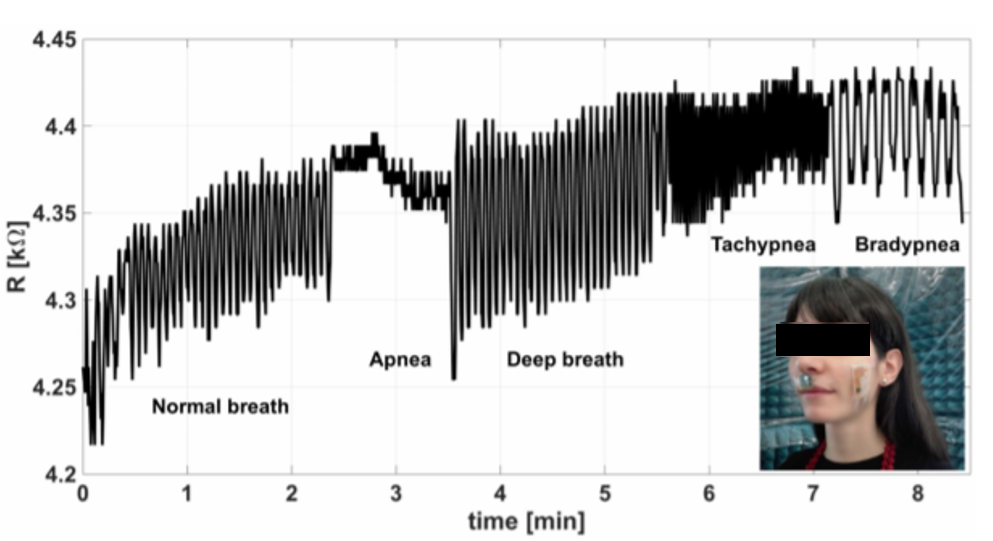}
\par\end{centering}
\begin{centering}
(c)
\par\end{centering}
\caption{RFID sensors for \emph{direct} respiratory function assessment. (a)
RFID breath humidity sensors for integration into a facemask (left)
and over the skin (right) (adapted from \cite{Caccami18,Caccami18breathanomalies}).
(b) Different prototypes of RFID skin-attachable breath temperature
sensors (adapted from \cite{Occhiuzzi18}). (c) Recording of a respiratory
pattern (adapted from \cite{Caccami18breathanomalies}). \label{fig:resp_cough}}
\end{figure}

\subsubsection{Breath}

RFID devices for breath monitoring can allow the avoidance of intrusive
nasal probes and chest bands \cite{Massaroni19}.

The first category of devices relies on \emph{direct} measurement
of the human breath through RFID temperature or moisture sensors to
be attached under the nose or over the facemask. A flexible RFID sensor
was demonstrated in \cite{Caccami18breathanomalies} and \cite{Caccami18}
(Fig. \ref{fig:resp_cough}.a) for integration into a facemask or
directly stuck on the face and provided with a graphene-oxide electrode
to monitor the moisture emitted during breathing. This sensor is able
to detect the inhalation/exhalation cycles and abnormal patterns of
respiration like apnea, tachypnea, etc. (Fig. \ref{fig:resp_cough}.c).
Simpler and lower-cost breath detection \cite{Occhiuzzi18} can only
involve a temperature sensor (even integrated into the RFID IC itself).
In this case, the overall device is smaller and suitable to be directly
applied close to the nostrils as an epidermal patch (Fig. \ref{fig:resp_cough}.b).
It measures the temperature gradient of the air flux that is correlated
to the inhalation and exhalation rhythm and depth. To obtain better
accuracy, bilateral breath measurements can be achieved by dual-chip
sensor tags as in \cite{Miozzi21giorgia}.

\begin{figure}[t]
\begin{centering}
\includegraphics[width=8.8cm]{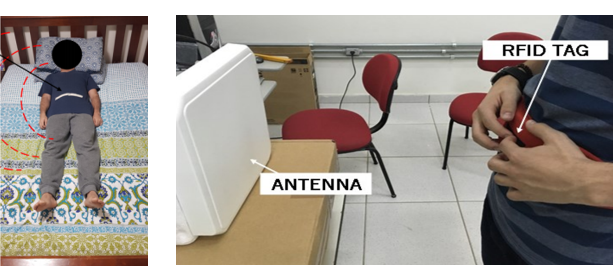}
\par\end{centering}
\begin{raggedright}
~~~~~~~~(a)~~~~~~~~~~~~~~~~~~~~~~~~~~~~~~~~~
(b)
\par\end{raggedright}
\begin{centering}
\includegraphics[width=8.8cm]{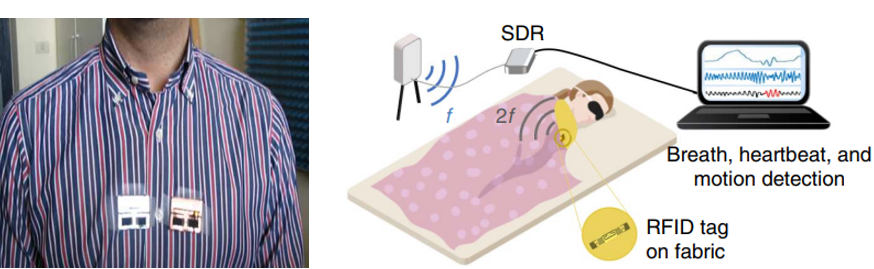}
\par\end{centering}
\begin{raggedright}
\textbf{~~~~~~~~~~~}(c)~~~~~~~~~~~~~~~~~~~~~~~~~~~~~~~~~~(d)
\par\end{raggedright}
\begin{centering}
\includegraphics[width=8.8cm]{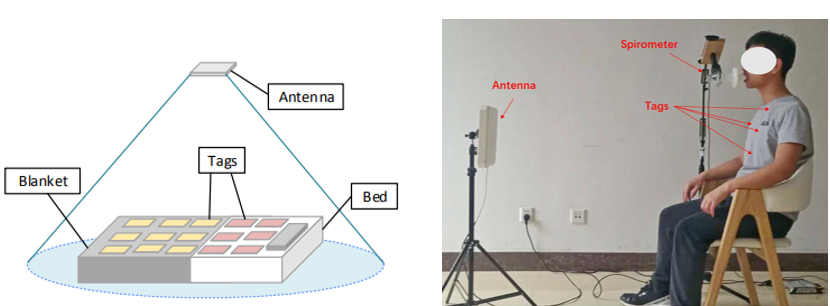}
\par\end{centering}
~~~~~~~~~~~~~~~~(e)~~~~~~~~~~~~~~~~~~~~~~~~~~~~~~~~~~~(f)

\caption{RFID sensors for \emph{indirect} respiratory function assessment.
(a) Sleep respiration monitoring system with passive COTS RFID tags
(adapted from \cite{Hussain19}). (b) RSSI-based respiratory monitoring
system with passive RFID tags (adapted from \cite{Araujo18}). (c)
Motion inertial based cough monitoring RFID sensor (adapted from \cite{Manzari12}).
(d) Fabric-integrated RFID tag for respiratory assessment during sleep
(adapted from \cite{Sharma18}). (e) Non-invasive sleep monitoring
system based on bed-integrated RFID (adapted from \cite{Hu17}). (f)
Measurement setup of the RF-RVM system (adapted from \cite{Chang21}).\label{fig:breath_sens}}
\end{figure}
\begin{figure}[t]
\begin{centering}
\includegraphics[width=8.5cm]{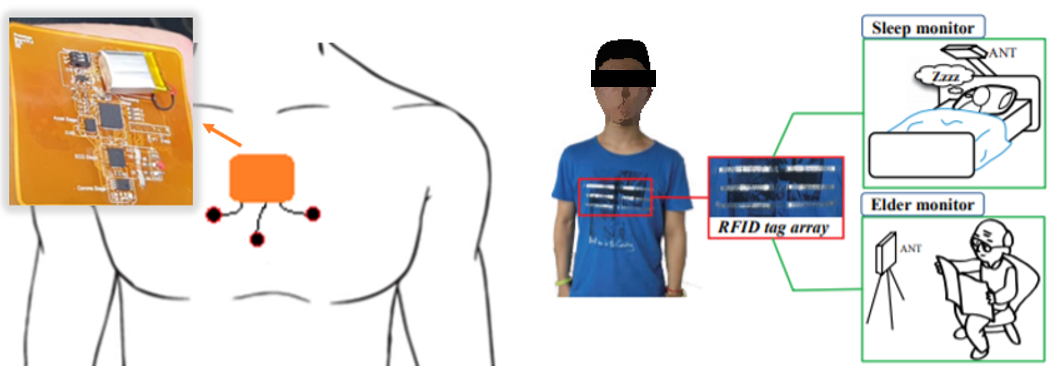}
\par\end{centering}
\begin{raggedright}
~~~~~~~~~~~~~~~~(a) ~~~~~~~~~~~~~~~~~~~~~~~~~~~~~~~~(b)
\par\end{raggedright}
\begin{centering}
\includegraphics[scale=0.5]{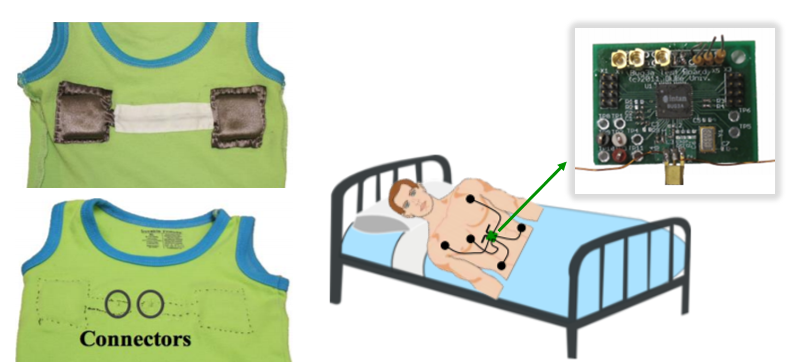}
\par\end{centering}
\begin{raggedright}
~~~~~~~~~~~~~~(c) ~~~~~~~~~~~~~~~~~~~~~~~~~~~~~~(d)
\par\end{raggedright}
\caption{RFID sensors for cardiovascular evaluation. (a) Epidermal ECG RFID
device (adapted from \cite{Horne20}). (b) RF-ECG system with an array
of COTS RFID tags (adapted from \cite{Wang18}). (c) Fabric-integrated
hearth rate RFID device (adapted from \cite{Agezo16}). (d) Battery-free
multichannel ECG sensor (adapted from \cite{Besnoff13}).\label{fig:cardio}}
\end{figure}

An \emph{indirect} breath monitoring avoiding sensors on the face
can be instead based on the measurement of the torso expansion and
contraction by placing regular RFID tags on the abdomen, or even on
the bed \cite{Hu17}, for overnight monitoring. Breath identification
is based on the measurements of the variations of RSSI \cite{Hussain19,Araujo18,Hui18}
and phase \cite{Wang19,Sharma18,Chang21}. Sleep respiration rate
and sleep apnea can be detected through a fixed reader antenna at
the bedside, and RFID tags can be attached \cite{Hussain19} (Fig.
\ref{fig:breath_sens}.a) or fabric-integrated \cite{Sharma18} (Fig.
\ref{fig:breath_sens}.d) at the abdomen level. For instance, the
\emph{TagBreathe} platform in \cite{Wang19} resorts to an array of
COTS RFID tags to enhance the measurement robustness. Similarly, the
respiratory volume can be retrieved by collecting the temporal phase
information from tags attached to the chest and abdomen \cite{Chang21}
(Fig. \ref{fig:breath_sens}.f). Moreover, by placing the RFID tag
on the abdomen, the respiratory rate in a medical examination-like
situation \cite{Araujo18} (Fig. \ref{fig:breath_sens}.b) can be
evaluated too.

By adding other RFID tags, for example, at the wrist area, additional
physiological parameters can be simultaneously collected, such as
heart rate, blood pressure, and respiration rate \cite{Hui18}.

\subsubsection{Oxygen Saturation}

The oxygen saturation in the blood can be captured by a wearable RFID
device \cite{Wang17oximetry} comprising a reflective oxygen probe
and the standard CMOS technology, whereas an active RFID tag is used
to store and stream the data.

\subsection{Cardiovascular Evaluation}

ECG data can be collected and streamed in real-time by means of an
ultra-low-power RFID BAP device to be placed at the center of the
chest, where standard clip electrodes collect the ECG signal \cite{Horne20,Wang13}
(Fig. \ref{fig:cardio}.a). The Heart Rate Variability (HRV) can be
quantified as well \cite{Wang18} with an array of COTS RFID tags
attached to the chest area within the clothes (Fig. \ref{fig:cardio}.b),
exploiting the processing of the RSSI as described above for breath.
Although the RFID reader captures the RF signal reflected from both
the heart movement due to heartbeat and the tag movement caused by
respiration, the estimated HRV is comparable to existing wired techniques.
Hearth Rate monitoring for infants can be achieved too by sewing an
ECG-enabled RFID sensor \cite{Vora15}, comprising fabric electrodes,
onto the front bodice of a baby onesie \cite{Agezo16} (Fig. \ref{fig:cardio}.c).
Multi-electrode RFID sensors (Fig. \ref{fig:cardio}.d) have been
experimented too, for a more accurate ECG recording without the use
of an on-board battery \cite{Besnoff13}.

\begin{figure}[t]
\begin{centering}
\includegraphics[scale=0.5]{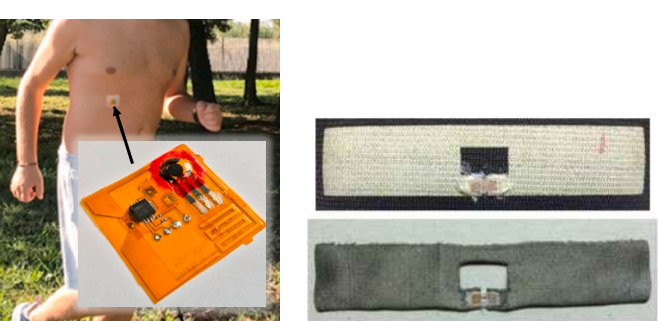}
\par\end{centering}
\begin{raggedright}
~~~~~~~~~~~~~~~~~~(a) ~~~~~~~~~~~~~~~~~~~~~~~~~(b)
\par\end{raggedright}
\caption{RFID sensors for sweat monitoring. (a) Epidermal RFID device for pH
sweat monitoring (adapted from \cite{Mazzaracchio20}). (b) Textile-integrated
RFID sweat rate sensor tags (adapted from \cite{Merilampi16}).\label{fig:sweat}}
\end{figure}

\begin{figure}[t]
\begin{centering}
\includegraphics[width=5.5cm]{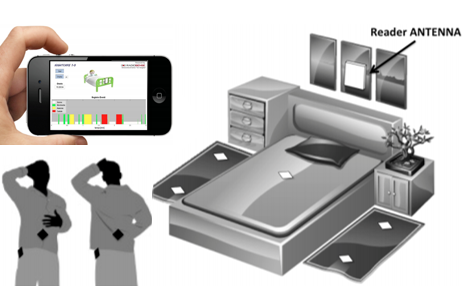}
\par\end{centering}
\begin{centering}
(a)
\par\end{centering}
\begin{centering}
\includegraphics[width=8.8cm]{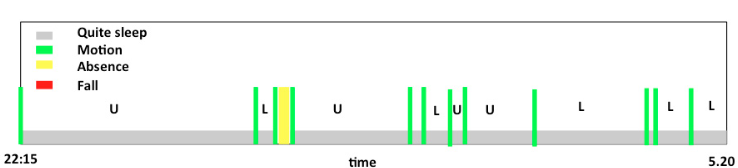}
\par\end{centering}
\begin{centering}
(b)
\par\end{centering}
\caption{(a)\emph{ NightCare} system for sleep monitoring (adapted from \cite{Occhiuzzi14},
\cite{Occhiuzzi14night}). (b) Classification of sleep quality and
different body postures (adapted from \cite{Occhiuzzi14night}).\label{fig:sleep}}
\end{figure}

\subsection{Psycho-Physical Stress Monitoring}

A pandemic may also impact the psychological sphere of patients and
operators \cite{Passavanti21}. For example, the onset of dehydration,
elevated biochemical stress, and increasing overnight sweat loss are
critical indicators associated with COVID-19 \cite{Daanen20}.

\subsubsection{Sweat Analysis}

Early signs or precursors of psychological diseases can be identified
by the chemical analysis of the sweat \cite{Sonner15}. The pH index,
which is correlated with the presence of altered electrolytes, can
be wirelessly monitored through a flexible RFID device (Fig. \ref{fig:sweat}.a)
\cite{Mazzaracchio20} equipped with an electrochemical printed iridium-oxide
sensor. Measurements can be taken on the fly in battery-less mode
from a distance up to $1$ m or continuously, provided that the device
works in battery-assisted mode \cite{Miozzi19}. Hence, monitoring
can be executed even during motion \cite{Nappi19}, as in the case
of nurse activity in a hospital. Sweat RFID sensors can also be integrated
into textiles (by carving electro-textiles or by screen-printing onto
regular threads, Fig. \ref{fig:sweat}.b), exploiting the change in
the tag's substrate when it absorbs sweat \cite{Merilampi16}.

\subsubsection{Quality of Sleep}

COVID-19, and in general emergency conditions, also affects the quality
of sleep. As suggested by \cite{AdansDester20}, measurements during
sleep might provide a significant insight on the health status, and
in some cases, it may be the only way to monitor a worker's health
effectively. For instance, the \emph{NightCare} system \cite{Occhiuzzi14,Occhiuzzi14night,Occhiuzzi10}
(Fig. \ref{fig:sleep}.a) can quantitatively evaluate the quality
and phenomenology of the sleep through a set of wearable passive tags
placed onto the sleeper\textquoteright s clothes and by also exploiting
ambient tags placed over the bed and onto the nearby carpets (Fig.
\ref{fig:sleep}.b). Finally, the sleeper can be provided with wearable
temperature sensor tags to additionally detect anomalous temperature
transients during the night \cite{AmendolaBovesecchi16}.

\section{Conclusions}

The presented analysis suggests that an RFID-based country-level infrastructure
could provide valuable support to the management of pandemics. The
availability of devices is rich and currently increasing. Research
outcomes from laboratories could quickly move to proof of concepts
and then to products.

When applied over the large scale to a pandemic-like scenario, RFID
frameworks are expected to generate an unprecedented amount of data.
Hence Edge Computing \cite{Buyya19} is needed to mitigate the data
stream, whereas Artificial Intelligence \cite{Nordlinger20} should
be used to extract profiles and identify anomalous events out of the
sea of data.

Furthermore, even more worrying issues for the social and industrial
acceptance of the RFID-centric approach to the pandemic war are (\emph{i})
the implementation costs of RFID systems and (\emph{ii}) the privacy.
Regarding the implementation, although the costs of RFID systems in
hospital is not negligible \cite{Coustasse2013}, the estimated return
of investment would be worth the initial cost \cite{Hakim2006,Smith2019}
since the same system can be used for multiple purposes (e.g. for
the tracking staff and equipment \cite{Cheng2016,Najera2011}, even
across different hospitals \cite{Meiller2011}). Then, concerns about
the privacy and safety related to the RFID labellig of humans can
lead to oral and practical forms of resistance, as the tampering or
the misuse of the systems \cite{Tarantini2019}. To preserve the anonymity
of users, a \emph{Crowd System} \cite{Lee2009} can be employed, wherein
the tags inside ``crowd zones'' repeatedly swap their identifiers.
To secure the contact tracing, RFID readers can exploit a blockchain
architecture through the internet \cite{Garg2020} so that a person
wearing the RFID tag can be notified if he/she has encountered a confirmed
infected person without disclosing his/her identity. Moreover, in
all the cases, the coverage of the reading infrastructure needs to
be optimized with ad-hoc network design methodologies \cite{Oztekin2010,Elewe2016,Panunzio21},
in compliance with the electromagnetic compatibility issues \cite{VanderTogt2011}.
Despite these open issues, studies on technology adoption predict
that the RFID will be increasingly accepted and widespread in healthcare
settings \cite{Chong2015}.

The main limitations of UHF RFID remain the limited read distance
and the need for a continuous remote power source to activate tags.
The latter problem could be mitigated through energy harvesting \cite{Huang18,Georgiadis14},
e.g., photovoltaic cells for environmental and on-object tags \cite{Abdelnour19},
or piezoelectric \cite{Shieh14}, vibration \cite{Khemmanee19} and
body-heat \cite{Jauregi17} powering for on-body sensors. Promising
opportunities would also arise by exploiting commodities signals to
implement backscattering communications among passive tags and routers,
access points and even smartphone and wearable devices. Finally, emerging
5G infrastructures \cite{Li18} will provide a further boost to improve
interoperability, coverage efficiency and bit-rate.

\section*{Acknowledgments}

The work was developed in the framework of the Dual-Skin project FISR
2020, founded by MIUR, Italy.

\bibliographystyle{IEEEtran}
\bibliography{RFID_COVID_29}

\begin{thebibliography}{100}
\providecommand{\url}[1]{#1}
\csname url@samestyle\endcsname
\providecommand{\newblock}{\relax}
\providecommand{\bibinfo}[2]{#2}
\providecommand{\BIBentrySTDinterwordspacing}{\spaceskip=0pt\relax}
\providecommand{\BIBentryALTinterwordstretchfactor}{4}
\providecommand{\BIBentryALTinterwordspacing}{\spaceskip=\fontdimen2\font plus
\BIBentryALTinterwordstretchfactor\fontdimen3\font minus
  \fontdimen4\font\relax}
\providecommand{\BIBforeignlanguage}[2]{{%
\expandafter\ifx\csname l@#1\endcsname\relax
\typeout{** WARNING: IEEEtran.bst: No hyphenation pattern has been}%
\typeout{** loaded for the language `#1'. Using the pattern for}%
\typeout{** the default language instead.}%
\else
\language=\csname l@#1\endcsname
\fi
#2}}
\providecommand{\BIBdecl}{\relax}
\BIBdecl

\bibitem{Arons20}
M.~Arons \emph{et~al.}, ``{Presymptomatic SARS-CoV-2 Infections and
  Transmission in a Skilled Nursing Facility},'' \emph{New England Journal of
  Medicine}, vol. 382, April 2020.

\bibitem{Hospitalization}
H.~Ritchie \emph{et~al.}, ``{Coronavirus (COVID-19) Hospitalizations},''
  \url{https://ourworldindata.org/covid-hospitalizations}, 2021.

\bibitem{Qian2020}
M.~Quian and J.~Jiang, ``{COVID-19} and social distancing,'' \emph{Journal of
  Public Health: From Theory to Practice}, vol. 2020, May 2020, doi:
  10.1007/s10389-020-01321-z.

\bibitem{Whitelaw20}
S.~Whitelaw, M.~Mamas, E.~Topol, and H.~Van~Spall, ``{Applications of digital
  technology in COVID-19 pandemic planning and response},'' \emph{The Lancet
  Digital Health}, June 2020.

\bibitem{WHOvaccines}
{World Health Organization}, ``{COVID-19 vaccination: supply and logistics
  guidance},''
  \url{https://www.who.int/publications/i/item/who-2019-ncov-vaccine-deployment-logistics-2021-1},
  2021.

\bibitem{vanDoremalen20}
N.~van Doremalen \emph{et~al.}, ``{Aerosol and surface stability of HCoV-19
  (SARS-CoV-2) compared to SARS-CoV-1},'' \emph{medRxiv}, 2020.

\bibitem{deMan20}
P.~de~Man, S.~Paltansing, D.~Ong, N.~Vaessen, G.~Nielen, and J.~Koeleman,
  ``{Outbreak of COVID-19 in a nursing home associated with aerosol
  transmission as a result of inadequate ventilation},'' \emph{Clinical
  infectious diseases: an official publication of the Infectious Diseases
  Society of America}, August 2020.

\bibitem{Ding20}
X.~Ding \emph{et~al.}, ``{Wearable Sensing and Telehealth Technology with
  Potential Applications in the Coronavirus Pandemic},'' \emph{IEEE Reviews in
  Biomedical Engineering}, vol.~PP, pp. 1--1, May 2020.

\bibitem{AdansDester20}
C.~P. Adans-Dester \emph{et~al.}, ``{Can mHealth Technology Help Mitigate the
  Effects of the COVID-19 Pandemic? - Supplementary materials},'' \emph{IEEE
  Open Journal of Engineering in Medicine and Biology}, vol.~1, pp. 243--248,
  2020.

\bibitem{Nasajpour20}
M.~Nasajpour, S.~Pouriyeh, R.~Parizi, M.~Dorodchi, M.~Valero, and H.~Arabnia,
  ``{Internet of Things for Current COVID-19 and Future Pandemics: an
  Exploratory Study},'' \emph{Journal of Healthcare Informatics Research},
  vol.~4, December 2020.

\bibitem{Dong21}
Y.~Dong and Y.~Yao, ``Iot platform for covid-19 prevention and control: A
  survey,'' \emph{IEEE Access}, vol.~9, pp. 49\,929--49\,941, 2021.

\bibitem{Habibzadeh20}
H.~{Habibzadeh}, K.~{Dinesh}, O.~{Rajabi Shishvan}, A.~{Boggio-Dandry},
  G.~{Sharma}, and T.~{Soyata}, ``{A Survey of Healthcare Internet of Things
  (HIoT): A Clinical Perspective},'' \emph{IEEE Internet of Things Journal},
  vol.~7, no.~1, pp. 53--71, 2020.

\bibitem{Occhiuzzi13sensorless}
C.~{Occhiuzzi}, S.~{Caizzone}, and G.~{Marrocco}, ``{Passive UHF RFID antennas
  for sensing applications: Principles, methods, and classifcations},''
  \emph{IEEE Antennas and Propagation Magazine}, vol.~55, no.~6, pp. 14--34,
  2013.

\bibitem{farsens}
Farsens, ``Farsens endless opportunities: wireless battery free sensors,''
  \url{http://www.farsens.com/en/}, 2020.

\bibitem{Mazzaracchio20}
V.~Mazzaracchio, L.~Fiore, S.~Nappi, G.~Marrocco, and F.~Arduini,
  ``{Medium-distance affordable, flexible and wireless epidermal sensor for pH
  monitoring in sweat},'' \emph{Talanta}, vol. 222, p. 121502, August 2020.

\bibitem{Yap16}
H.~Yap, B.~Ang, J.~Lim, J.~Goh, and C.-H. Yeow, ``A fabric-regulated soft
  robotic glove with user intent detection using {EMG and RFID} for hand
  assistive application,'' in \emph{Proceedings - IEEE International Conference
  on Robotics and Automation}, 2016, pp. 3537--3542.

\bibitem{Dobkin:2007}
D.~M. Dobkin, \emph{The RF in RFID: passive UHF RFID in Practice}.\hskip 1em
  plus 0.5em minus 0.4em\relax Amsterdam: Elsevier, 2007.

\bibitem{Occhiuzzi19industry}
C.~{Occhiuzzi}, S.~{Amendola}, S.~{Nappi}, N.~{D'Uva}, and G.~{Marrocco},
  ``{RFID Technology for Industry 4.0: Architectures and Challenges},'' in
  \emph{2019 IEEE International Conference on RFID Technology and Applications
  (RFID-TA)}, 2019, pp. 181--186.

\bibitem{WHO2020}
{World Health Organization}, ``{Rational use of personal protective equipment
  (PPE) for coronavirus disease (COVID-19)},''
  \url{https://www.who.int/publications/i/item/rational-use-of-personal-protective-equipment-for-coronavirus-disease-(covid-19)-and-considerations-during-severe-shortages},
  2020.

\bibitem{Lockhart2020}
S.~Lockhart, J.~Naidu, C.~Badh, and L.~Duggan, ``Simulation as a tool for
  assessing and evolving your current personal protective equipment: lessons
  learned during the coronavirus disease (covid-19) pandemic,'' \emph{Canadian
  Journal of Anesthesia}, vol.~67, no.~7, pp. 895--896, Mar 2020.

\bibitem{Zamora2006}
J.~Zamora, J.~Murdoch, B.~Simchison, and A.~Day, ``Contamination: A comparison
  of 2 personal protective systems,'' \emph{CMAJ}, vol. 175, no.~3, pp.
  249--254, Aug 2006.

\bibitem{Mostaghimi2020}
A.~Mostaghimi \emph{et~al.}, ``{Regulatory and Safety Considerations in
  Deploying a Locally Fabricated, Reusable Face Shield in a Hospital Responding
  to the COVID-19 Pandemic},'' \emph{Clinical and Translational Resource and
  Technology Insights}, vol.~1, no.~1, pp. 139--151, Dec 2020.

\bibitem{Alexandre2017}
D.~Alexandre, M.~Prieto, F.~Beaumont, R.~Taiar, and G.~Polidori, ``Wearing lead
  aprons in surgical operating rooms: ergonomic injuries evidenced by infrared
  thermography,'' \emph{Journal of Surgical Research}, vol. 209, pp. 227--233,
  Mar 2017.

\bibitem{RaabeVanessa2012}
N.~Raabe~Vanessa and B.~Matthias, ``Infection control during filoviral
  hemorrhagic fever outbreaks,'' \emph{Journal of Global Infectious Diseases},
  vol.~4, no.~1, pp. 69--74, Mar 2012.

\bibitem{Occhiuzzi_design_2013}
C.~Occhiuzzi and G.~Marrocco, ``Constrained-design of passive {UHF RFID} sensor
  antennas,'' \emph{IEEE Transactions on Antennas and Propagation}, vol.~61,
  no.~6, pp. 2972--2980, 2013.

\bibitem{OcchiuzziAccuracy16}
------, ``Precision and accuracy in {UHF-RFID} power measurements for passive
  sensing,'' \emph{IEEE Sensors Journal}, vol.~16, no.~9, pp. 3091--3098, 2016.

\bibitem{QU2011}
X.~Qu, L.~T. Simpson, and P.~Stanfield, ``A model for quantifying the value of
  {RFID-enabled} equipment tracking in hospitals,'' \emph{Advanced Engineering
  Informatics}, vol.~25, no.~1, pp. 23 -- 31, 2011.

\bibitem{Mun2007}
I.~Mun, A.~Kantrowitz, P.~Carmel, K.~Mason, and D.~Engels, ``Active {RFID}
  system augmented with {2D} barcode for asset management in a hospital
  setting,'' in \emph{2007 IEEE International Conference on RFID, IEEE RFID
  2007}, Grapevine, TX, USA, Mar 2007, pp. 205--211.

\bibitem{Hakim2006}
H.~Hakim, R.~Renouf, and J.~Enderle, ``Passive {RFID} asset monitoring system
  in hospital environments,'' in \emph{Bioengineering, Proceedings of the
  Northeast Conference}, vol. 2006, Easton, PA, USA, Apr 2006, pp. 217--218.

\bibitem{Jarritt2014}
P.~Jarritt, ``Managing mobile medical devices using radiofrequency
  identification,'' \emph{British Journal of Health Care Management}, vol.~20,
  no.~1, pp. 16--21, 2014.

\bibitem{Meiller2011}
Y.~Meiller, S.~Bureau, W.~Zhou, and S.~Piramuthu, ``{RFID}-embedded decision
  support for tracking surgical equipment,'' in \emph{Proceedings of the Annual
  Hawaii International Conference on System Sciences}, Koloa, HI, United
  States, Jan 2011.

\bibitem{Ostbye2003}
T.~{\O}stbye \emph{et~al.}, ``Evaluation of an infrared/radiofrequency
  equipment-tracking system in a tertiary care hospital,'' \emph{Journal of
  Medical Systems}, vol.~27, no.~4, pp. 367--380, 2003.

\bibitem{Shirehjini2012}
A.~Shirehjini, A.~Yassine, and S.~Shirmohammadi, ``Equipment location in
  hospitals using {RFID}-based positioning system,'' \emph{IEEE Transactions on
  Information Technology in Biomedicine}, vol.~16, no.~6, pp. 1058--1069, 2012.

\bibitem{Tsai2019}
M.-H. Tsai, C.-S. Pan, C.-W. Wang, J.-M. Chen, and C.-B. Kuo, ``{RFID} medical
  equipment tracking system based on a location-based service technique,''
  \emph{Journal of Medical and Biological Engineering}, vol.~39, no.~1, pp.
  163--169, 2019.

\bibitem{Kumar2008}
S.~Kumar, E.~Swanson, and T.~Tran, ``{RFID in the healthcare supply chain:
  usage and application},'' \emph{International Journal of Health Care Quality
  Assurance}, vol.~22, no.~1, pp. 67--81, Apr 2008.

\bibitem{Coustasse2013}
A.~Coustasse, S.~Tomblin, and C.~Slack, ``{Impact of radio-frequency
  identification ({RFID}) technologies on the hospital supply chain: a
  literature review},'' \emph{Perspectives in Health Information Management},
  pp. 1--17, Oct 2013.

\bibitem{Smith2019}
A.~D. Smith, ``{RFID Applications in Healthcare Systems From an Operational
  Perspective},'' \emph{International Journal of Systems and Society}, vol.~6,
  no.~2, pp. 1--28, Nov 2019.

\bibitem{Borelli2013}
G.~Borelli, P.~Orr\`{u}, and F.~Zedda, ``Performance analysis of a healthcare
  supply chain. a {RFID} system implementation design,'' in \emph{Proceedings
  of the Summer School Francesco Turco}, Senigallia, Italy, Sept 2013, pp.
  42--47.

\bibitem{Katsiri20162}
E.~Katsiri, K.~Pramatari, A.~Billiris, A.~Kaiafas, A.~Christodoulakis, and
  H.~Karanikas, ``{RFcure}: An {RFID} based blood bank/healthcare information
  management system,'' in \emph{IFMBE Proceedings}, Paphos, Cyprus, Sept 2016,
  pp. 853--858.

\bibitem{Kalra2012}
R.~Kalra, P.~Shetty, S.~Mutalik, U.~Nayak, M.~Reddy, and N.~Udupa,
  ``Pharmaceutical applications of radio-frequency identification,''
  \emph{Systematic Reviews in Pharmacy}, vol.~3, no.~1, pp. 24--30, Feb 2012.

\bibitem{Cole2008}
P.~H. Cole and D.~C. Ranasinghe, \emph{{Networked RFID Systems and Lightweight
  Cryptography }}, Sept 2008.

\bibitem{Inaba2008}
T.~Inaba, \emph{EPC system for a safe and secure supply chain and how it is
  applied}, 2008.

\bibitem{SAFKHANI2020}
M.~Safkhani, S.~Rostampour, Y.~Bendavid, and N.~Bagheri, ``{IoT} in medical and
  pharmaceutical: Designing lightweight {RFID} security protocols for ensuring
  supply chain integrity,'' \emph{Computer Networks}, vol. 181, p. 107558,
  2020.

\bibitem{Huang2010}
G.~Huang, Z.~Qin, T.~Qu, and Q.~Dai, ``{RFID}-enabled pharmaceutical regulatory
  traceability system,'' in \emph{Proceedings of 2010 IEEE International
  Conference on RFID-Technology and Applications, RFID-TA 2010}, 2010, pp.
  211--216.

\bibitem{Raj2019}
R.~Raj, N.~Rai, and S.~Agarwal, ``{Anticounterfeiting in Pharmaceutical Supply
  Chain by establishing Proof of Ownership},'' in \emph{TENCON 2019 - 2019 IEEE
  Region 10 Conference (TENCON)}, Kochi, India, Oct 2019, pp. 1572--1577.

\bibitem{Toyoda2017}
K.~{Toyoda}, P.~T. {Mathiopoulos}, I.~{Sasase}, and T.~{Ohtsuki}, ``A novel
  blockchain-based product ownership management system (poms) for
  anti-counterfeits in the post supply chain,'' \emph{IEEE Access}, vol.~5, pp.
  17\,465--17\,477, June 2017.

\bibitem{Bauk2018}
S.~Bauk, A.~Schmeink, and J.~Colomer, ``{An RFID model for improving workers'
  safety at the seaport in transitional environment},'' \emph{Transport},
  vol.~33, no.~2, pp. 353--363, 2018.

\bibitem{KELM2013}
A.~Kelm \emph{et~al.}, ``Mobile passive radio frequency identification ({RFID})
  portal for automated and rapid control of personal protective equipment
  ({PPE}) on construction sites,'' \emph{Automation in Construction}, vol.~36,
  pp. 38 -- 52, 2013.

\bibitem{Musu2014}
C.~{Musu}, V.~{Popescu}, and D.~{Giusto}, ``Workplace safety monitoring using
  {RFID} sensors,'' in \emph{2014 22nd Telecommunications Forum Telfor
  (TELFOR)}, Belgrade, Serbia, Nov 2014, pp. 656--659.

\bibitem{Mahmad2016}
M.~K.~N. Mahmad \emph{et~al.}, ``Applications of radio frequency identification
  ({RFID}) in mining industries,'' in \emph{{International Conference on
  Innovative Research 2016}}, Iasi, Romania, May 2016.

\bibitem{Hung2019}
M.~H. Hung, L.~T. Lan, and H.~S. Hong, ``{A Deep Learning-Based Method for
  Real-Time Personal Protective Equipment Detection},'' \emph{Journal of
  Science and Technique}, vol. 199, no.~13, pp. 23--34, June 2019.

\bibitem{Pradana2019}
R.~D.~W. {Pradana} \emph{et~al.}, ``Identification system of personal
  protective equipment using convolutional neural network (cnn) method,'' in
  \emph{2019 International Symposium on Electronics and Smart Devices (ISESD)},
  Badung-Bali, Indonesia, Oct 2019, pp. 1--6.

\bibitem{Manfred2012}
M.~Helmus, A.~Kelm, A.~Meins-Becker, D.~Platz, and K.~M. Javad, ``Life cycle
  data of {PPE with RFID} sensors - new research results,'' in \emph{{11th
  European Seminar on Personal Protective Equipment}}, Saariselka, Finland, Jan
  2012, pp. 1--5.

\bibitem{Sole2013}
M.~{Sole}, C.~{Musu}, F.~{Boi}, D.~{Giusto}, and V.~{Popescu}, ``{RFID} sensor
  network for workplace safety management,'' in \emph{2013 IEEE 18th Conference
  on Emerging Technologies Factory Automation (ETFA)}, Cagliari, Italy, Sept
  2013, pp. 1--4.

\bibitem{BARROTORRES2012}
S.~Barro-Torres, T.~M. Fern\'andez-Caram\'es, H.~J. P\'erez-Iglesias, and C.~J.
  Escudero, ``Real-time personal protective equipment monitoring system,''
  \emph{Computer Communications}, vol.~36, no.~1, pp. 42 -- 50, 2012.

\bibitem{Mandar2020}
E.~M. Mandar, W.~Dachry, and B.~Bensassi, ``Toward a real-time personal
  protective equipment compliance control system based on {RFID} technology,''
  in \emph{{1st International Conference of Advanced Computing and
  Informatics}}, Casablanca, Morocco, Apr 2020, pp. 553--565.

\bibitem{Cabreira2015}
C.~M. {Cabreira} \emph{et~al.}, ``{RFID} applied to protective equipment
  inspection,'' in \emph{2015 IEEE Brasil RFID}, S\~{a}o Paulo, Brasil, Oct
  2015, pp. 1--4.

\bibitem{Smith2005}
J.~Smith \emph{et~al.}, ``{RFID}-based techniques for human-activity
  detection,'' \emph{Communications of the ACM}, vol.~48, no.~9, pp. 39--44,
  2005.

\bibitem{DONG2018}
S.~Dong, H.~Li, and Q.~Yin, ``Building information modeling in combination with
  real time location systems and sensors for safety performance enhancement,''
  \emph{Safety Science}, vol. 102, pp. 226 -- 237, 2018.

\bibitem{Bianco21}
G.~M. Bianco and G.~Marrocco, ``Sensorized facemask with moisture-sensitive
  {RFID} antenna,'' \emph{IEEE Sensors Letters}, vol.~5, no.~3, pp. 1--4, Feb.
  2021, doi:10.1109/LSENS.2021.3059348.

\bibitem{Liu2017}
H.~Liu and Z.~Yao, ``The research on recycling management model of medical
  waste based on {RFID} technology,'' \emph{Fresenius Environmental Bulletin},
  vol.~26, no.~8, pp. 5069--5081, Aug 2017.

\bibitem{Liu2020}
H.~Liu, Z.~Yao, F.~Chang, and S.~Meyer, ``An {RFID}-based medical waste
  transportation management system: Assessment of a new model on a hospital in
  china,'' \emph{Fresenius Environmental Bulletin}, vol.~29, no.~2, pp.
  773--784, Feb 2020.

\bibitem{Katsiri2016}
E.~Katsiri and K.~Moschou, ``A pervasive computing system for the remote
  management of hospital waste,'' \emph{Journal of Communications Software and
  Systems}, vol.~12, no.~1, pp. 53--66, Mar 2016.

\bibitem{Nolz2014}
P.~Nolz, N.~Absi, and D.~Feillet, ``A stochastic inventory routing problem for
  infectious medical waste collection,'' \emph{Networks}, vol.~63, no.~1, pp.
  82--95, Jan 2014.

\bibitem{Sun2019}
S.~Sun, J.~Hu, Y.~Cao, and W.~Zhou, ``Discussion on the application of {RFID}
  technology in medical waste management,'' in \emph{Proceedings - 10th
  International Conference on Information Technology in Medicine and Education,
  ITME 2019}, Qingdao, China, Aug 2019, pp. 120--124.

\bibitem{Bose2011}
I.~Bose and S.~Yan, ``The green potential of {RFID} projects: A case-based
  analysis,'' \emph{IT Professional}, vol.~13, no.~1, pp. 41 -- 47, Jan. 2011.

\bibitem{Liu20172}
H.~Liu and Z.~Yao, ``Research on the reverse logistics management of medical
  waste based on the {RFID} technology,'' \emph{Fresenius Environmental
  Bulletin}, vol.~25, no.~12, pp. 8084--8092, Dec. 2017.

\bibitem{Gonzalez21}
F.~Villa-Gonzalez, R.~Bhattacharyya, and S.~Sarma, ``Single and bulk
  identification of plastics in the recycling chain using chipless {RFID}
  tags,'' in \emph{15th Annual Conference on RFID}, 2021.

\bibitem{Bonanni2005}
L.~Bonanni, E.~Arroyo, C.-H. Lee, and T.~Selker, ``Smart sinks: Real-world
  opportunities for context-aware interaction,'' in \emph{{Conference on Human
  Factors in Computing Systems}}, Portland, OR, USA, April 2005, pp.
  1232--1235.

\bibitem{Do2009}
E.-L. Do, ``Technological interventions for hand hygiene adherence: Research
  and intervention for smart patient room,'' in \emph{Joining Languages,
  Cultures and Visions - CAADFutures 2009, Proceedings of the 13th
  International CAAD Futures Conference}, Montr\'{e}al, Canada, June 2009, pp.
  303--313.

\bibitem{Radhakrishna2015}
K.~Radhakrishna \emph{et~al.}, ``Real-time feedback for improving compliance to
  hand sanitization among healthcare workers in an open layout icu using
  radiofrequency identification,'' \emph{Journal of Medical Systems}, vol.~39,
  no.~6, 2015.

\bibitem{Decker2016}
A.~Decker, G.~Cipriano, G.~Tsouri, and J.~Lavigne, ``Monitoring pharmacy
  student adherence to world health organization hand hygiene indications using
  radio frequency identification,'' \emph{American Journal of Pharmaceutical
  Education}, vol.~80, no.~3, 2016.

\bibitem{Johnson2012}
R.~Johnson, G.~Tsouri, and E.~Walsh, ``Continuous and automated measuring of
  compliance of hand-hygiene procedures using finite state machines and
  {RFID},'' \emph{IEEE Instrumentation and Measurement Magazine}, vol.~15,
  no.~2, pp. 8--12, 2012.

\bibitem{Meydanci2013}
M.~Meydanci, C.~Adali, M.~Ertas, M.~Dizbay, and A.~Akan, ``{RFID} based hand
  hygiene compliance monitoring station,'' in \emph{2013 IEEE International
  Conference on Control System, Computing and Engineering (ICCSCE 2013)},
  Penang, Malaysia, Nov 2013, pp. 573--576.

\bibitem{Pineles2014}
L.~Pineles \emph{et~al.}, ``Accuracy of a radiofrequency identification
  ({RFID}) badge system to monitor hand hygiene behavior during routine
  clinical activities,'' \emph{American Journal of Infection Control}, vol.~42,
  no.~2, pp. 144--147, 2014.

\bibitem{Pletersek2012}
A.~Pleter\v{s}ek, M.~Sok, and J.~Trontelj, ``Monitoring, control and
  diagnostics using {RFID} infrastructure,'' \emph{Journal of Medical Systems},
  vol.~36, no.~6, pp. 3733--3739, 2012.

\bibitem{onlineAXZON}
Axzon, ``{RFM3300-E Magnus-S3 M3E passive sensor IC},''
  \url{https://axzon.com/rfm3300-e-magnus-s3-m3e-passive-sensor-ic/}.

\bibitem{onlineSL900}
AMS, ``{SL900A EPC Gen2 Sensor Tag and Data Logger IC},''
  \url{https://ams.com/documents/20143/36005/SL900A_DS000294_5-00.pdf/d399f354-b0b6-146f-6e98-b124826bd737}.

\bibitem{onlineEM4325}
{EM Microelectronic EM4325},
  \url{https://www.emmicroelectronic.com/sites/default/files/products/datasheets/4325-ds_0.pdf}.

\bibitem{onlineAsygn}
Asygn, ``{Asygn AS321X - Passive UHF RFID chip},''
  \url{https://asygn.com/as321x/}.

\bibitem{Amendola14}
S.~{Amendola}, R.~{Lodato}, S.~{Manzari}, C.~{Occhiuzzi}, and G.~{Marrocco},
  ``{RFID Technology for IoT-Based Personal Healthcare in Smart Spaces},''
  \emph{IEEE Internet of Things Journal}, vol.~1, no.~2, pp. 144--152, 2014.

\bibitem{WHO2020_2}
{World Health Organization}, ``{Personal protective equipment (PPE) for
  different healthcare activities},''
  \url{https://www.who.int/bangladesh/emergencies/coronavirus-disease-(covid-19)-update/steps-to-put-on-personal-protective-equipment-(ppe)},
  2020.

\bibitem{Holland2020}
M.~Holland, D.~J. Zaloga, and C.~S. Friderici, ``{COVID-19 Personal Protective
  Equipment (PPE) for the emergency physician},'' \emph{Visual Journal of
  Emergency Medicine}, vol.~19, 2020.

\bibitem{Ippolito2020}
M.~Ippolito, C.~Gregoriano, A.~Cortegiani, and P.~Iozzo, ``{Counterfeit
  filtering facepiece respirators are posing an additional risk to health care
  workers during COVID-19 pandemic},'' \emph{American Journal of Infection and
  Control}, vol.~48, pp. 853--858, 2020.

\bibitem{Pitts2020}
P.~Pitts, ``The spreading cancer of counterfeit drugs,'' \emph{Journal of
  Commercial Biotechnology}, vol.~25, no.~3, pp. 20--14, 2020.

\bibitem{Cook2020}
T.~M. Cook, ``Personal protective equipment during the coronavirus disease
  (covid) 2019 pandemic - a narrative review,'' \emph{Anaesthesia}, vol.~75,
  no.~7, pp. 920--927, Apr 2020.

\bibitem{Houghton2020}
C.~Houghton \emph{et~al.}, ``{Barriers and facilitators to healthcare workers'
  adherence with infection prevention and control (IPC) guidelines for
  respiratory infectious diseases: a rapid qualitative evidence synthesis},''
  \emph{Cochrane Database of Systematic Reviews}, no.~4, 2020.

\bibitem{Hignett2020}
S.~Highnett, R.~Welsh, and J.~Banerjee, ``{Human factors issues of working in
  personal protective equipment during the COVID-19 pandemic},''
  \emph{Anaesthesia}, 2020.

\bibitem{Jiang2020}
Q.~Jiang \emph{et~al.}, ``The prevalence, characteristics, and related factors
  of pressure injury in medical staff wearing personal protective equipment
  against covid-19 in china: A multicentre cross-sectional survey,''
  \emph{International Wound Journal}, vol.~17, no.~5, pp. 1300--1309, Apr 2020.

\bibitem{Gao2016}
S.~Gao \emph{et~al.}, ``{Performance of N95 FFRs Against Combustion and NaCl
  Aerosols in Dry and Moderately Humid Air: Manikin-based Study},''
  \emph{Annals of Occupational Hygiene}, vol.~60, no.~6, pp. 748--760, Apr
  2016.

\bibitem{Pacitto2019}
A.~Pacitto \emph{et~al.}, ``{Effectiveness of commercial face masks to reduce
  personal PM exposure},'' \emph{Science of Total Environment}, vol. 650,
  no.~1, pp. 1592--1590, Feb. 2019.

\bibitem{Klemes2020}
J.~J. Klemes, Y.~V. Fang, and P.~Jian, ``{The energy and environmental
  footprints of COVID-19 fighting measures e PPE, disinfection, supply
  chains},'' \emph{Energy}, vol. 211, August 2020.

\bibitem{Bardram2011}
J.~E. {Bardram}, A.~{Doryab}, R.~M. {Jensen}, P.~M. {Lange}, K.~L.~G.
  {Nielsen}, and S.~T. {Petersen}, ``Phase recognition during surgical
  procedures using embedded and body-worn sensors,'' in \emph{2011 IEEE
  International Conference on Pervasive Computing and Communications (PerCom)},
  Seattle, WA, USA, Mar 2011, pp. 45--53.

\bibitem{Denis2019}
S.~Denis, R.~Berkvens, and M.~Weyn, ``A survey on detection, tracking and
  identification in radio frequency-based device-free localization,''
  \emph{Sensors (Switzerland)}, vol.~19, no.~23, 2019.

\bibitem{Gholamhosseini2019}
L.~Gholamhosseini, F.~Sadoughi, and A.~Safaei, ``Hospital real-time location
  system (a practical approach in healthcare): A narrative review article,''
  \emph{Iranian Journal of Public Health}, vol.~48, no.~4, pp. 593--602, 2019.

\bibitem{Vivaldi2020}
F.~Vivaldi \emph{et~al.}, ``A temperature-sensitive {RFID} tag for the
  identification of cold chain failures,'' \emph{Sensors and Actuators, A:
  Physical}, vol. 313, pp. 1--18, 2020.

\bibitem{Hu2020}
L.~Hu, C.~Xiang, and C.~Qi, ``Research on traceability of cold chain logistics
  based on {RFID} and {EPC},'' in \emph{IOP Conference Series: Materials
  Science and Engineering}, vol. 790, no.~1, 2020, pp. 1--5.

\bibitem{Caizzone2011}
S.~Caizzone, C.~Occhiuzzi, and G.~Marrocco, ``Multi-chip {RFID} antenna
  integrating shape-memory alloys for detection of thermal thresholds,''
  \emph{IEEE Transactions on Antennas and Propagation}, vol.~59, no.~7, pp.
  2488--2494, Jul 2011.

\bibitem{Mahase2020}
\BIBentryALTinterwordspacing
E.~Mahase, ``Covid-19: What do we know about the late stage vaccine
  candidates?'' \emph{The BMJ}, vol. 371, pp. 1--2, Nov 2020. [Online].
  Available: \url{https://www.bmj.com/content/371/bmj.m4576}
\BIBentrySTDinterwordspacing

\bibitem{WHO2020ter}
{World Health Organization}, ``Advice on the use of masks in the context of
  {COVID-19},''
  \url{https://www.who.int/publications/i/item/advice-on-the-use-of-masks-in-the-community-during-home-care-and-in-healthcare-settings-in-the-context-of-the-novel-coronavirus-(2019-ncov)-outbreak},
  2020.

\bibitem{Caccami2017}
M.~C. {Caccami}, M.~Y.~S. {Mulla}, C.~{Di Natale}, and G.~{Marrocco},
  ``Wireless monitoring of breath by means of a graphene oxide-based
  radiofrequency identification wearable sensor,'' in \emph{2017 11th European
  Conference on Antennas and Propagation (EUCAP)}, 2017, pp. 3394--3396.

\bibitem{WINDFELD2015}
\BIBentryALTinterwordspacing
E.~S. Windfeld and M.~S.-L. Brooks, ``Medical waste management - a review,''
  \emph{Journal of Environmental Management}, vol. 163, pp. 98--108, Nov 2015.
  [Online]. Available:
  \url{http://www.sciencedirect.com/science/article/pii/S0301479715302176}
\BIBentrySTDinterwordspacing

\bibitem{HANNAN2015}
M.~Hannan, M.~{Abdulla Al Mamun}, A.~Hussain, H.~Basri, and R.~Begum, ``A
  review on technologies and their usage in solid waste monitoring and
  management systems: Issues and challenges,'' \emph{Waste Management},
  vol.~43, pp. 509--523, 2015.

\bibitem{Wu10}
D.~{Wu}, W.~W.~Y. {Ng}, P.~P.~K. {Chan}, H.~{Ding}, B.~{Jing}, and D.~S.
  {Yeung}, ``{Access control by RFID and face recognition based on neural
  network},'' in \emph{2010 International Conference on Machine Learning and
  Cybernetics}, vol.~2, 2010, pp. 675--680.

\bibitem{Pan12}
X.~{Pan}, ``{Research and Implementation of Access Control System Based on RFID
  and FNN-Face Recognition},'' in \emph{2012 Second International Conference on
  Intelligent System Design and Engineering Application}, 2012, pp. 716--719.

\bibitem{Louati12}
T.~{Louati}, F.~{Ounnar}, P.~{Pujo}, and C.~{Pistoresi}, ``{A simulation test
  bed of intelligent access control based on biometric and RFID},'' in
  \emph{2012 IEEE International Conference on RFID-Technologies and
  Applications (RFID-TA)}, 2012, pp. 297--302.

\bibitem{Shafin15}
M.~Kishwar~Shafin \emph{et~al.}, ``{Development of an RFID based access control
  system in the context of Bangladesh},'' in \emph{2015 International
  Conference on Innovations in Information, Embedded and Communication Systems
  (ICIIECS)}, 2015, pp. 1--5.

\bibitem{Bakht19}
K.~{Bakht}, A.~U. {Din}, A.~{Shehzadi}, and M.~{Aftab}, ``{Design of an
  Efficient Authentication and Access Control System Using RFID},'' in
  \emph{2019 3rd International Conference on Energy Conservation and Efficiency
  (ICECE)}, 2019, pp. 1--4.

\bibitem{Amendola_Sept17_SCISSOR}
S.~{Amendola}, C.~{Occhiuzzi}, and G.~{Marrocco}, ``{RFID sensing networks for
  critical infrastructure security: A real testbed in an energy smart grid},''
  in \emph{2017 IEEE International Conference on RFID Technology Application
  (RFID-TA)}, 2017, pp. 106--110.

\bibitem{Manzari12}
S.~{Manzari}, S.~{Pettinari}, and G.~{Marrocco}, ``{Miniaturized and tunable
  wearable RFID tag for body-centric applications},'' in \emph{2012 IEEE
  International Conference on RFID-Technologies and Applications (RFID-TA)},
  2012, pp. 239--243.

\bibitem{Buffi19}
A.~{Buffi}, B.~{Tellini}, A.~{Motroni}, and P.~{Nepa}, ``{A Phase-based Method
  for UHF RFID Gate Access Control},'' in \emph{2019 IEEE International
  Conference on RFID Technology and Applications (RFID-TA)}, 2019, pp.
  131--135.

\bibitem{Wen16}
S.~{Wen}, H.~{Heidari}, A.~{Vilouras}, and R.~{Dahiya}, ``{A wearable
  fabric-based {RFID} skin temperature monitoring patch},'' in \emph{2016 IEEE
  SENSORS}, 2016, pp. 1--3.

\bibitem{AmendolaBovesecchi16}
S.~Amendola, G.~Bovesecchi, A.~Palombi, P.~Coppa, and G.~Marrocco, ``{Design,
  Calibration and Experimentation of an Epidermal RFID Sensor for Remote
  Temperature Monitoring},'' \emph{IEEE Sensors Journal}, vol.~16, pp. 1--1,
  October 2016.

\bibitem{iTek}
{i-Tek - Body Temperature Monitoring System for Employees and Civil Services},
  \url{https://infoteksoftware.com/rfid-solution/rfid-based-body-temperature-monitoring-for-workers.html?vertical=health-care-vertical}.

\bibitem{Chen2009}
Y.-W. Chen \emph{et~al.}, ``A {RFID} model of transferring and tracking trauma
  patients after a large disaster,'' in \emph{2009 IEEE/INFORMS International
  Conference on Service Operations, Logistics and Informatics, SOLI 2009},
  Chicago, IL, USA, Jul 2009, pp. 98--101.

\bibitem{Chen20092}
P.-J. Chen, Y.-F. Chen, S.-K. Chai, and Y.-F. Huang, ``Implementation of an
  {RFID}-based management system for operation room,'' in \emph{2009
  International Conference on Machine Learning and Cybernetics}, vol.~5,
  Baoding, China, Jul 2009, pp. 2933--2938.

\bibitem{Chowdhury2007}
B.~Chowdhury and R.~Khosla, ``{RFID}-based hospital real-time patient
  management system,'' in \emph{Proceedings - 6th IEEE/ACIS International
  Conference on Computer and Information Science, ICIS 2007; 1st IEEE/ACIS
  International Workshop on e-Activity, IWEA 2007}, Melbourne, Australia, Jul
  2007, pp. 363--368.

\bibitem{Harry2014}
T.~Harry, M.~Taylor, R.~Fletcher, A.~Mundt, and T.~Pawlicki, ``Passive tracking
  of linac clinical flow using radiofrequency identification technology,''
  \emph{Practical Radiation Oncology}, vol.~4, no.~1, pp. e85--e90, 2014.

\bibitem{Perez2018}
M.~P\'{e}rez, C.~Dafonte, and A.~G\'{o}mez, ``Traceability in patient
  healthcare through the integration of {RFID} technology in an {ICU} in a
  hospital,'' \emph{Sensors (Switzerland)}, vol.~18, no.~5, 2018.

\bibitem{Cao2014}
Q.~Cao, D.~R. Jones, and H.~Sheng, ``Contained nomadic information
  environments: Technology, organization, and environment influences on
  adoption of hospital {RFID} patient tracking,'' \emph{Information and
  Management}, vol.~51, no.~2, pp. 225 -- 239, 2014.

\bibitem{Cheng2016}
C.-H. Cheng and Y.-H. Kuo, ``{RFID} analytics for hospital ward management,''
  \emph{Flexible Services and Manufacturing Journal}, vol.~28, no.~4, pp.
  593--616, 2016.

\bibitem{Chhetri2019}
B.~Chhetri, A.~Alsadoon, P.~Prasad, H.~Venkata, and A.~Elchouemi, ``Enhanced
  weighted centroid localization in {RFID} technology: Patient movement
  tracking in hospital,'' in \emph{2019 5th International Conference on
  Advanced Computing and Communication Systems, ICACCS 2019}, Coimbatore,
  India, Mar 2019, pp. 910--915.

\bibitem{Iadanza2008}
E.~Iadanza, F.~Dori, R.~Miniati, and R.~Bonaiuti, ``Patients tracking and
  identifying inside hospital: A multilayer method to plan an {RFId}
  solution,'' in \emph{Proceedings of the 30th Annual International Conference
  of the IEEE Engineering in Medicine and Biology Society, EMBS'08 -
  "Personalized Healthcare through Technology"}, 2008, pp. 1462--1465.

\bibitem{Magliulo2012}
M.~{Magliulo}, L.~{Cella}, and R.~{Pacelli}, ``Novel technique radio frequency
  identification ({RFID}) based to manage patient flow in a radiotherapy
  department,'' in \emph{Proceedings of 2012 IEEE-EMBS International Conference
  on Biomedical and Health Informatics}, 2012, pp. 972--975.

\bibitem{Mapa2015}
L.~Mapa and K.~Saha, ``Application of {RFID} technology in patient management
  system,'' in \emph{{Proceedings of 122nd ASEE Annual Conference and
  Exposition}}, Seattle, WA, USA, June 2015, pp. 2--17.

\bibitem{Najera2011}
P.~Najera, J.~Lopez, and R.~Roman, ``Real-time location and inpatient care
  systems based on passive {RFID},'' \emph{Journal of Network and Computer
  Applications}, vol.~34, no.~3, pp. 980--989, 2011.

\bibitem{Kim2008}
D.-S. Kim, J.~Kim, S.-H. Kim, and S.~Yoo, ``Design of {RFID} based the patient
  management and tracking system in hospital,'' in \emph{Proceedings of the
  30th Annual International Conference of the IEEE Engineering in Medicine and
  Biology Society, EMBS'08 - "Personalized Healthcare through Technology"},
  2008, pp. 1459--1461.

\bibitem{Newman-Casey2020}
P.~Newman-Casey, J.~Musser, L.~Niziol, K.~Shedden, D.~Burke, and A.~Cohn,
  ``Designing and validating a low-cost real time locating system to
  continuously assess patient wait times,'' \emph{Journal of Biomedical
  Informatics}, vol. 106, 2020.

\bibitem{Kapoor21}
A.~Kapoor \emph{et~al.}, ``Estimating physical work-load on ed clinicians and
  staff using real-time location systems,'' in \emph{15th Annual Conference on
  RFID}, Apr. 2021.

\bibitem{Perez2017}
M.~P\'{e}rez, G.~Gonz\'{a}lez, and C.~Dafonte, ``The development of an {RFID}
  solution to facilitate the traceability of patient and pharmaceutical data,''
  \emph{Sensors (Switzerland)}, vol.~17, no.~10, 2017.

\bibitem{Bahri2007}
S.~Bahri, ``Enhancing quality of data through automated {SARS} contact tracing
  method using {RFID} technology,'' \emph{International Journal of Networking
  and Virtual Organisations}, vol.~4, no.~2, pp. 145--162, May 2007.

\bibitem{Chang2011}
Y.-T. Chang, S.~Syed-Abdul, C.-Y. Tsai, and Y.-C. Li, ``A novel method for
  inferring {RFID} tag reader recordings into clinical events,''
  \emph{International Journal of Medical Informatics}, vol.~80, no.~12, pp.
  872--880, Dec. 2011.

\bibitem{Garg2020}
L.~{Garg}, E.~{Chukwu}, N.~{Nasser}, C.~{Chakraborty}, and G.~{Garg},
  ``Anonymity preserving iot-based covid-19 and other infectious disease
  contact tracing model,'' \emph{IEEE Access}, vol.~8, pp. 159\,402--159\,414,
  Aug. 2020.

\bibitem{North2013}
D.~Lowery-North \emph{et~al.}, ``Measuring social contacts in the emergency
  department,'' \emph{PLoS ONE}, vol.~8, no.~8, Aug. 2013, art. no. e70854.

\bibitem{Cattuto2010}
C.~Cattuto, W.~van~den Broeck, A.~Barrat, V.~Colizza, J.-F. Pinton, and
  A.~Vespignani, ``Dynamics of person-to-person interactions from distributed
  {RFID} sensor networks,'' \emph{PLoS ONE}, vol.~5, no.~7, 2010, art. no.
  e11596.

\bibitem{Stehle2011}
J.~Stehl\'{e} \emph{et~al.}, ``Simulation of an {SEIR} infectious disease model
  on the dynamic contact network of conference attendees,'' \emph{BMC
  Medicine}, vol.~9, 2011, art. no. 87.

\bibitem{Smieszek2016}
T.~Smieszek, S.~Castell, A.~Barrat, C.~Cattuto, P.~White, and G.~Krause,
  ``Contact diaries versus wearable proximity sensors in measuring contact
  patterns at a conference: Method comparison and participants' attitudes,''
  \emph{BMC Infectious Diseases}, vol.~16, no.~1, 2016, art. no. 341.

\bibitem{Nikitin2012}
P.~Nikitin, S.~Ramamurthy, R.~Martinez, and K.~Rao, ``Passive tag-to-tag
  communication,'' in \emph{2012 IEEE International Conference on RFID, RFID
  2012}, 2012, pp. 177--184.

\bibitem{Marrocco2012}
G.~Marrocco and S.~Caizzone, ``Electromagnetic models for passive tag-to-tag
  communications,'' \emph{IEEE Transactions on Antennas and Propagation},
  vol.~60, no.~11, pp. 5381--5389, 2012.

\bibitem{Karimi2017}
Y.~Karimi, A.~Athalye, S.~Das, P.~Djuric, and M.~Stanacevic, ``Design of a
  backscatter-based tag-to-tag system,'' in \emph{2017 IEEE International
  Conference on RFID, RFID 2017}, 2017, pp. 6--12.

\bibitem{DeDonno2011}
D.~De~Donno, F.~Ricciato, L.~Catarinucci, and L.~Tarricone, ``Design and
  applications of a software-defined listener for {UHF RFID} systems,'' in
  \emph{IEEE MTT-S International Microwave Symposium Digest}, 2011.

\bibitem{Piumwardane21}
D.~Piumwardane, C.~Rohner, and T.~Voigt, ``Reliable flooding in dense
  backscatter-based tag-to-tag networks,'' in \emph{15th Annual Conference on
  RFID}, 2021.

\bibitem{Borisenko2013}
A.~Borisenko, M.~Bolic, and M.~Rostamian, ``Intercepting {UHF RFID} signals
  through synchronous detection,'' \emph{Eurasip Journal on Wireless
  Communications and Networking}, vol. 2013, no.~1, 2013.

\bibitem{Bolic2015}
M.~Bolic, M.~Rostamian, and P.~Djuric, ``Proximity detection with {RFID}: A
  step toward the internet of things,'' \emph{IEEE Pervasive Computing},
  vol.~14, no.~2, pp. 70--76, 2015.

\bibitem{Isella2011}
L.~Isella \emph{et~al.}, ``Close encounters in a pediatric ward: Measuring
  face-to-face proximity and mixing patterns with wearable sensors,''
  \emph{PLoS ONE}, vol.~6, no.~2, 2011.

\bibitem{Saab2016}
S.~Saab and H.~Msheik, ``Novel {RFID}-based pose estimation using single
  stationary antenna,'' \emph{IEEE Transactions on Industrial Electronics},
  vol.~63, no.~3, pp. 1842--1852, 2016.

\bibitem{AlvarezNarciandi2018}
G.~Alvarez-Narciandi, J.~Laviada, M.~Pino, and F.~Las-Heras, ``Attitude
  estimation based on arrays of passive {RFID} tags,'' \emph{IEEE Transactions
  on Antennas and Propagation}, vol.~66, no.~5, pp. 2534--2544, 2018.

\bibitem{Krigslund2011}
R.~Krigslund, P.~Popovski, G.~Pedersen, and K.~Bank, ``Potential of {RFID}
  systems to detect object orientation,'' in \emph{IEEE International
  Conference on Communications}, 2011.

\bibitem{Barbot2020}
N.~Barbot, O.~Rance, and E.~Perret, ``Angle sensor based on chipless {RFID}
  tag,'' \emph{IEEE Antennas and Wireless Propagation Letters}, vol.~19, no.~2,
  pp. 233--237, 2020.

\bibitem{Akbar2019}
M.~Akbar, D.~Taylor, and G.~Durgin, ``Planar position and orientation
  estimation using a 5.8 {GHz RFID} system,'' in \emph{2019 IEEE International
  Conference on RFID Technology and Applications, RFID-TA 2019}, 2019, pp.
  252--257.

\bibitem{Xiao2018}
F.~Xiao, Z.~Wang, N.~Ye, R.~Wang, and X.-Y. Li, ``One more tag enables
  fine-grained {RFID} localization and tracking,'' \emph{IEEE/ACM Transactions
  on Networking}, vol.~26, no.~1, pp. 161--174, 2018.

\bibitem{Cho05}
N.~Cho, S.-J. Song, S.~Kim, S.~Kim, and H.-J. Yoo, ``{A 5.1-/spl mu/W UHF RFID
  tag chip integrated with sensors for wireless environmental monitoring},'' in
  \emph{Proceedings of the 31st European Solid-State Circuits Conference, 2005.
  ESSCIRC 2005.}, 2005, pp. 279--282.

\bibitem{Qi14}
Z.~Qi, Y.~Zhuang, X.~Li, W.~Liu, Y.~Du, and B.~Wang, ``{Full passive UHF RFID
  Tag with an ultra-low power, small area, high resolution temperature sensor
  suitable for environment monitoring},'' \emph{Microelectronics Journal},
  vol.~45, no.~1, pp. 126--131, 2014.

\bibitem{Occhiuzzi_Feb17_SCISSOR}
C.~Occhiuzzi, S.~Amendola, S.~Manzari, S.~Caizzone, and G.~Marrocco,
  ``{Configurable RFID Sensing Antenna Breadboard for Industrial IoT},''
  \emph{Electronics Letters}, vol.~53, December 2016.

\bibitem{Oprea08}
I.~A. Oprea, N.~Barsan, U.~Weimar, M.-L. Bauersfeld, D.~Ebling, and
  J.~W\"{o}llenstein, ``{Capacitive Humidity Sensors on Flexible RFID
  Labels},'' \emph{Sensors and Actuators B: Chemical}, vol. 132, pp. 404--410,
  June 2008.

\bibitem{Siden09}
J.~{Siden}, J.~{Gao}, and B.~{Neubauer}, ``{Microstrip antennas for remote
  moisture sensing using passive RFID},'' in \emph{2009 Asia Pacific Microwave
  Conference}, 2009, pp. 2375--2378.

\bibitem{Virtanen11}
J.~{Virtanen}, L.~{Ukkonen}, T.~{Bjorninen}, A.~Z. {Elsherbeni}, and
  L.~{Syd\"{a}nheimo}, ``{Inkjet-Printed Humidity Sensor for Passive UHF RFID
  Systems},'' \emph{IEEE Transactions on Instrumentation and Measurement},
  vol.~60, no.~8, pp. 2768--2777, 2011.

\bibitem{Manzari12_humidity}
S.~{Manzari}, C.~{Occhiuzzi}, S.~{Nawale}, A.~{Catini}, C.~{Di Natale}, and
  G.~{Marrocco}, ``{Humidity Sensing by Polymer-Loaded UHF RFID Antennas},''
  \emph{IEEE Sensors Journal}, vol.~12, no.~9, pp. 2851--2858, 2012.

\bibitem{Amin13}
Y.~Amin, Y.~Feng, Q.~Chen, L.-R. Zheng, and H.~Tenhunen, ``{RFID antenna
  humidity sensor co-design for USN applications},'' \emph{IEICE Electronics
  Express}, vol.~10, pp. 20\,130\,003--20\,130\,003, January 2013.

\bibitem{Manzari13}
S.~{Manzari}, A.~{Catini}, C.~{Di Natale}, and G.~{Marrocco}, ``{Ambient
  sensing by chemical-loaded UHF-RFIDs},'' in \emph{2013 7th European
  Conference on Antennas and Propagation (EuCAP)}, 2013, pp. 1718--1720.

\bibitem{Manzari14_J}
S.~{Manzari} and G.~{Marrocco}, ``{Modeling and Applications of a
  Chemical-Loaded UHF RFID Sensing Antenna With Tuning Capability},''
  \emph{IEEE Transactions on Antennas and Propagation}, vol.~62, no.~1, pp.
  94--101, 2014.

\bibitem{Colella16}
R.~{Colella}, L.~{Catarinucci}, and L.~{Tarricone}, ``{Improved Battery-Less
  Augmented RFID Tag: Application on Ambient Sensing and Control},'' \emph{IEEE
  Sensors Journal}, vol.~16, no.~10, pp. 3484--3485, 2016.

\bibitem{WHO2009}
{World Health Organization}, ``{Hand Hygiene: Why, How \& When?}''
  \url{https://www.who.int/gpsc/5may/Hand_Hygiene_Why_How_and_When_Brochure.pdf},
  2020.

\bibitem{Dawson2014}
C.~Dawson and J.~Mackrill, ``Review of technologies available to improve hand
  hygiene compliance - are they fit for purpose?'' \emph{Journal of Infection
  Prevention}, vol.~15, no.~6, pp. 222--228, 2014.

\bibitem{Chapter_SCISSOR}
S.~Amendola, C.~Occhiuzzi, S.~Manzari, and G.~Marrocco, ``{RFID}-based
  multi-level sensing network for industrial internet of things,'' in
  \emph{Studies in Computational Intelligence}, June 2018, pp. 1--24.

\bibitem{KRETZSCHMAR2020}
M.~E. Kretzschmar, G.~Rozhnova, M.~C.~J. Bootsma, M.~{van Boven}, J.~H. H.~M.
  {van de Wijgert}, and M.~J.~M. Bonten, ``Impact of delays on effectiveness of
  contact tracing strategies for {COVID-19}: a modelling study,'' \emph{The
  Lancet Public Health}, vol.~5, no.~8, pp. e452--e459, 2020.

\bibitem{Mareiniss2020}
D.~Mareiniss, ``The impending storm: Covid-19, pandemics and our overwhelmed
  emergency departments,'' \emph{American Journal of Emergency Medicine},
  vol.~38, no.~6, pp. 1293--1294, 2020.

\bibitem{Maringe2020}
C.~Maringe \emph{et~al.}, ``The impact of the {COVID-19} pandemic on cancer
  deaths due to delays in diagnosis in england, uk: a national,
  population-based, modelling study,'' \emph{The Lancet Oncology}, vol.~21,
  no.~8, pp. 1023--1034, 2020.

\bibitem{Makoni2020}
M.~Makoni, ``Covid-19 worsens zimbabwe's health crisis,'' \emph{Lancet (London,
  England)}, vol. 396, no. 10249, p. 457, 2020.

\bibitem{Spena20}
A.~Spena, L.~Palombi, M.~Corcione, M.~Carestia, and V.~Spena, ``{On the Optimal
  Indoor Air Conditions for SARS-CoV-2 Inactivation. An Enthalpy-Based
  Approach},'' \emph{International Journal of Environmental Research and Public
  Health}, vol.~17, p. 6083, August 2020.

\bibitem{Quraishi20}
S.~A. Quraishi, L.~Berra, and A.~Nozari, ``{Indoor temperature and relative
  humidity in hospitals: workplace considerations during the novel coronavirus
  pandemic},'' \emph{Occupational and Environmental Medicine}, vol.~77, no.~7,
  pp. 508--508, 2020.

\bibitem{IndoorCapacity}
``{COVID-19 Indoor Capacity Limitations: Instructions and Resources},''
  \url{https://www.sccgov.org/sites/covid19/Pages/covid-capacity.aspx}, 2021.

\bibitem{CDCfever}
{Centers for Disease Control and Prevention}, ``{Definitions of Symptoms for
  Reportable Illnesses},''
  \url{https://www.cdc.gov/quarantine/air/reporting-deaths-illness/definitions-symptoms-reportable-illnesses.html},
  2020.

\bibitem{Marrocco_Oct16_SCISSOR}
C.~{Occhiuzzi}, S.~{Amendola}, S.~{Manzari}, and G.~{Marrocco}, ``{Industrial
  RFID sensing networks for critical infrastructure security},'' in \emph{2016
  46th European Microwave Conference (EuMC)}, 2016, pp. 1335--1338.

\bibitem{Merilampi16}
S.~{Merilampi}, {Han He}, L.~{Syd\"{a}nheimo}, L.~{Ukkonen}, and J.~{Virkki},
  ``{The possibilities of passive UHF RFID textile tags as comfortable wearable
  sweat rate sensors},'' in \emph{2016 Progress in Electromagnetic Research
  Symposium (PIERS)}, 2016, pp. 3984--3987.

\bibitem{Bjorninen15}
T.~{Bjorninen}, J.~{Virkki}, L.~{Sydanheimo}, and L.~{Ukkonen},
  ``{Manufacturing of antennas for passive UHF RFID tags by direct write
  dispensing of copper and silver inks on textiles},'' in \emph{2015
  International Conference on Electromagnetics in Advanced Applications
  (ICEAA)}, 2015, pp. 589--592.

\bibitem{BookChapterAmendola}
S.~Amendola \emph{et~al.}, ``{UHF} epidermal sensors: Technology and
  applications,'' 11 2020, pp. 133--161.

\bibitem{Wang06}
S.-W. Wang, W.-H. Chen, C.-S. Ong, L.~Liu, and Y.-W. Chuang, ``{RFID
  Application in Hospitals: A Case Study on a Demonstration RFID Project in a
  Taiwan Hospital},'' in \emph{Proceedings of the 39th Annual Hawaii
  International Conference on System Sciences (HICSS'06)}, vol.~8, 2006, pp.
  184a--184a.

\bibitem{VanPraet2020}
J.~Van~Praet, B.~Claeys, A.-S. Coene, K.~Flor\'{e}, and M.~Reynders,
  ``Prevention of nosocomial {COVID-19}: another challenge of the pandemic,''
  \emph{Infection Control and Hospital Epidemiology}, vol.~41, no.~11, pp.
  1355--1356, Nov. 2020.

\bibitem{ECDC2020}
{European centre for disese prevention and control},
  \url{https://www.ecdc.europa.eu/sites/default/files/documents/covid-19-contact-tracing-public-health-management-third-update.pdf},
  2020.

\bibitem{Camera20}
F.~Camera, C.~Miozzi, F.~Amato, C.~Occhiuzzi, and G.~Marrocco, ``{Experimental
  Assessment of Wireless Monitoring of Axilla Temperature by Means of Epidermal
  Battery-Less RFID Sensors},'' \emph{IEEE Sensors Letters}, vol.~4, no.~11,
  pp. 1--4, 2020.

\bibitem{Occhiuzzi20}
C.~Occhiuzzi, S.~Parrella, F.~Camera, S.~Nappi, and G.~Marrocco, ``{RFID-based
  Dual-Chip Epidermal Sensing Platform for Human Skin Monitoring},'' \emph{IEEE
  Sensors Letters}, 2020.

\bibitem{Miozzi20thin}
C.~Miozzi, F.~Amato, and G.~Marrocco, ``{Performance and Durability of Thread
  Antennas as Stretchable Epidermal UHF RFID Tags},'' \emph{IEEE Journal of
  Radio Frequency Identification}, vol.~4, no.~4, pp. 398--405, 2020.

\bibitem{Camera19}
F.~{Camera} \emph{et~al.}, ``{Monitoring of temperature stress during
  firefighters training by means of RFID epidermal sensors},'' in \emph{2019
  IEEE International Conference on RFID Technology and Applications (RFID-TA)},
  2019, pp. 499--504.

\bibitem{Miozzi19}
C.~{Miozzi}, S.~{Nappi}, S.~{Amendola}, C.~{Occhiuzzi}, and G.~{Marrocco}, ``{A
  General-Purpose Configurable RFID Epidermal Board With a Two-Way Discrete
  Impedance Tuning},'' \emph{IEEE Antennas and Wireless Propagation Letters},
  vol.~18, no.~4, pp. 684--687, 2019.

\bibitem{Miozzi18}
C.~{Miozzi}, S.~{Nappi}, S.~{Amendola}, and G.~{Marrocco}, ``{A General-Purpose
  Small RFID Epidermal Datalogger for Continuous Human Skin Monitoring in
  Mobility},'' in \emph{2018 IEEE/MTT-S International Microwave Symposium -
  IMS}, 2018, pp. 371--373.

\bibitem{Miozzi18trial}
C.~Miozzi, S.~Amendola, A.~Bergamini, and G.~Marrocco, ``{Clinical Trial of
  Wireless Epidermal Temperature Sensors: preliminary results},'' in
  \emph{Nordic-Baltic Conference on Biomedical Engineering and Medical
  Physics}, June 2018, pp. 1041--1044.

\bibitem{Miozzi17}
C.~{Miozzi}, S.~{Amendola}, A.~{Bergamini}, and G.~{Marrocco}, ``{Reliability
  of a re-usable wireless Epidermal temperature sensor in real conditions},''
  in \emph{2017 IEEE 14th International Conference on Wearable and Implantable
  Body Sensor Networks (BSN)}, 2017, pp. 95--98.

\bibitem{Adame16}
T.~Adame, A.~Bel, A.~Carreras, J.~Melia-Segui, M.~Oliver, and R.~Pous,
  ``{CUIDATS: An RFID-WSN hybrid monitoring system for smart health care
  environments},'' \emph{Future Generation Computer Systems}, vol.~78, December
  2016.

\bibitem{Milici14}
S.~Milici, S.~Amendola, A.~Bianco, and G.~Marrocco, ``{Epidermal RFID Passive
  Sensor for Body Temperature Measurements},'' in \emph{2014 IEEE RFID
  Technology and Applications Conference, RFID-TA 2014}, October 2014, pp.
  140--144.

\bibitem{Chang21}
X.~{Chang}, J.~{Dai}, Z.~{Zhang}, K.~{Zhu}, and G.~{Xing}, ``{RF-RVM:
  Continuous Respiratory Volume Monitoring With COTS RFID Tags},'' \emph{IEEE
  Internet of Things Journal}, pp. 1--1, 2021.

\bibitem{Miozzi21giorgia}
C.~Miozzi, G.~Stendardo, G.~M. Bianco, F.~Montecchia, and G.~Marrocco,
  ``Dual-chip {RFID} on-skin tag for bilateral breath monitoring,'' in
  \emph{15th Annual Conference on RFID}, Apr. 2021.

\bibitem{Oliveira20}
V.~Oliveira, L.~Duarte, G.~Costa, M.~Macedo, and T.~Silveira, ``{Automation
  System for Six-minute Walk Test Using RFID Technology},'' in
  \emph{International Symposium on Networks, Computers and Communications},
  2020.

\bibitem{Wang19}
Y.~Wang and Y.~Zheng, ``{TagBreathe: Monitor Breathing with Commodity RFID
  Systems},'' \emph{IEEE Transactions on Mobile Computing}, vol.~PP, pp. 1--1,
  February 2019.

\bibitem{Hussain19}
Z.~Hussain, S.~Sagar, W.~E. Zhang, and Q.~Sheng, ``{A cost-effective and
  non-invasive system for sleep and vital signs monitoring using passive RFID
  tags},'' in \emph{MobiQuitous '19: Proceedings of the 16th EAI International
  Conference on Mobile and Ubiquitous Systems: Computing, Networking and
  Services}, November 2019, pp. 153--161.

\bibitem{Caccami18breathanomalies}
M.~C. Caccami, Y.~Mulla, C.~Occhiuzzi, C.~Natale, and G.~Marrocco, ``{Design
  and Experimentation of a Batteryless On-skin RFID Graphene-Oxide Sensor for
  the Monitoring and Discrimination of Breath Anomalies},'' \emph{IEEE Sensors
  Journal}, vol.~PP, pp. 1--1, August 2018.

\bibitem{Occhiuzzi18}
C.~{Occhiuzzi}, C.~{Caccami}, S.~{Amendola}, and G.~{Marrocco},
  ``{Breath-monitoring by means of Epidermal Temperature RFID Sensors},'' in
  \emph{2018 3rd International Conference on Smart and Sustainable Technologies
  (SpliTech)}, 2018, pp. 1--4.

\bibitem{Caccami18}
M.~C. {Caccami}, M.~Y.~S. {Mulla}, C.~{Di Natale}, and G.~{Marrocco},
  ``{Graphene oxide-based radiofrequency identification wearable sensor for
  breath monitoring},'' \emph{IET Microwaves, Antennas Propagation}, vol.~12,
  no.~4, pp. 467--471, 2018.

\bibitem{Araujo18}
J.~Araujo \emph{et~al.}, ``Passive {RFID} tag for respiratory frequency
  monitoring,'' in \emph{3rd International Symposium on Instrumentation
  Systems, Circuits and Transducers}, August 2018, pp. 1--5.

\bibitem{Hui18}
X.~Hui and E.~Kan, ``{Monitoring vital signs over multiplexed radio by
  near-field coherent sensing},'' \emph{Nature Electronics}, vol.~1, January
  2018.

\bibitem{Sharma18}
P.~{Sharma} and E.~C. {Kan}, ``{Sleep Scoring with a UHF RFID Tag by Near Field
  Coherent Sensing},'' in \emph{2018 IEEE/MTT-S International Microwave
  Symposium - IMS}, 2018, pp. 1419--1422.

\bibitem{Caccami17}
M.~C. {Caccami}, C.~{Miozzi}, M.~Y.~S. {Mulla}, C.~{Di Natale}, and
  G.~{Marrocco}, ``{An epidermal graphene oxide-based RFID sensor for the
  wireless analysis of human breath},'' in \emph{2017 IEEE International
  Conference on RFID Technology Application (RFID-TA)}, 2017, pp. 191--195.

\bibitem{Hou17}
Y.~{Hou}, Y.~{Wang}, and Y.~{Zheng}, ``{TagBreathe: Monitor Breathing with
  Commodity RFID Systems},'' in \emph{2017 IEEE 37th International Conference
  on Distributed Computing Systems (ICDCS)}, 2017, pp. 404--413.

\bibitem{Hu17}
X.~{Hu}, K.~{Naya}, P.~{Li}, T.~{Miyazaki}, and K.~{Wang}, ``{Non-invasive
  sleep monitoring based on RFID},'' in \emph{2017 IEEE 19th International
  Conference on e-Health Networking, Applications and Services (Healthcom)},
  2017, pp. 1--3.

\bibitem{Wang17oximetry}
Y.~Wang, ``{Design of the Pulse Oximetry Measurement Circuit and Its Sensing
  System Based On CMOS},'' \emph{IOSR Journal of Electrical and Electronics
  Engineering}, vol.~12, pp. 64--70, March 2017.

\bibitem{Horne20}
R.~Horne, J.~Batchelor, P.~Taylor, E.~Balaban, and A.~Casson, ``{Ultra-Low
  Power on Skin ECG using RFID Communication},'' in \emph{IEEE International
  Conference on Flexible and Printable Sensors and Systems}, 2020.

\bibitem{Wang18}
C.~Wang, L.~Xie, W.~Wang, Y.~Chen, Y.~Bu, and S.~Lu, ``{RF-ECG: Heart Rate
  Variability Assessment Based on COTS RFID Tag Array},'' \emph{Proceedings of
  the ACM on Interactive, Mobile, Wearable and Ubiquitous Technologies},
  vol.~2, pp. 1--26, July 2018.

\bibitem{Agezo16}
S.~Agezo, Y.~Zhang, Z.~Ye, S.~Chopra, S.~Vora, and T.~Kurzweg, ``{Battery-free
  RFID heart rate monitoring system},'' in \emph{2016 IEEE Wireless Health
  (WH)}, 2016, pp. 1--7.

\bibitem{Vora15}
S.~Vora, K.~Dandekar, and T.~Kurzweg, ``{Passive RFID tag based heart rate
  monitoring from an ECG signal},'' in \emph{Annual International Conference of
  the IEEE Engineering in Medicine and Biology Society}, vol. 2015, August
  2015, pp. 4403--4406.

\bibitem{Wang13}
C.-R. Wang, S.-Y. Lee, and W.-C. Lai, ``{An RFID tag system-on-chip with
  wireless ECG monitoring for intelligent healthcare systems},''
  \emph{Conference proceedings: Annual International Conference of the IEEE
  Engineering in Medicine and Biology Society. IEEE Engineering in Medicine and
  Biology Society. Conference}, vol. 2013, pp. 5489--5492, July 2013.

\bibitem{Besnoff13}
J.~S. {Besnoff}, T.~{Deyle}, R.~R. {Harrison}, and M.~S. {Reynolds},
  ``{Battery-free multichannel digital ECG biotelemetry using UHF RFID
  techniques},'' in \emph{2013 IEEE International Conference on RFID (RFID)},
  2013, pp. 16--22.

\bibitem{Nappi19}
S.~Nappi, V.~Mazzaracchio, L.~Fiore, F.~Arduini, and G.~Marrocco, ``{Flexible
  pH Sensor for Wireless Monitoring of the Human Skin from the Medimun
  Distances},'' in \emph{2019 IEEE International Conference on Flexible and
  Printable Sensors and Systems}, July 2019, pp. 1--3.

\bibitem{Occhiuzzi14}
C.~{Occhiuzzi}, C.~{Vallese}, S.~{Amendola}, S.~{Manzari}, and G.~{Marrocco},
  ``{Multi-channel processing of RFID backscattering for monitoring of
  overnight living},'' in \emph{2014 XXXIth URSI General Assembly and
  Scientific Symposium (URSI GASS)}, 2014, pp. 1--4.

\bibitem{Occhiuzzi14night}
C.~Occhiuzzi, C.~Vallese, S.~Amendola, S.~Manzari, and G.~Marrocco,
  ``{NIGHT-Care: A Passive RFID System for Remote Monitoring and Control of
  Overnight Living Environment},'' \emph{Procedia Computer Science}, vol.~32,
  pp. 190--197, December 2014.

\bibitem{Occhiuzzi10}
C.~{Occhiuzzi} and G.~{Marrocco}, ``{The RFID Technology for Neurosciences:
  Feasibility of Limbs' Monitoring in Sleep Diseases},'' \emph{IEEE
  Transactions on Information Technology in Biomedicine}, vol.~14, no.~1, pp.
  37--43, 2010.

\bibitem{online_ventilation}
{USA Today},
  \url{https://eu.usatoday.com/in-depth/graphics/2020/10/18/improving-indoor-air-quality-prevent-covid-19/3566978001/},
  2020.

\bibitem{CDC}
{Centers for Disease Control and Prevention}, ``{Symptoms of Coronavirus},''
  \url{https://www.cdc.gov/coronavirus/2019-ncov/symptoms-testing/symptoms.html},
  2020.

\bibitem{Petrilli20}
C.~Petrilli \emph{et~al.}, ``{Factors associated with hospital admission and
  critical illness among 5279 people with coronavirus disease 2019 in New York
  City: Prospective cohort study},'' \emph{BMJ}, vol. 369, p. m1966, May 2020.

\bibitem{Hasell20}
J.~Hasell, ``{Testing early, testing late: four countries' approaches to
  COVID-19 testing compared},''
  \url{https://ourworldindata.org/covid-testing-us-uk-korea-italy}, May 2020.

\bibitem{Sun20}
Q.~Sun, H.~Qiu, M.~Huang, and Y.~Yang, ``{Lower mortality of COVID-19 by early
  recognition and intervention: experience from Jiangsu Province},''
  \emph{Annals of Intensive Care}, vol.~10, 2020.

\bibitem{Caccami18epidermal}
M.~C. {Caccami}, C.~{Miozzi}, V.~{Greco}, and G.~{Marrocco}, ``{Epidermal
  radio-sensors for wireless detection of physiological parameters and sense
  augmentation},'' in \emph{12th European Conference on Antennas and
  Propagation (EuCAP 2018)}, 2018, pp. 1--5.

\bibitem{Panunzio21covid}
N.~Panunzio, G.~M. Bianco, C.~Occhiuzzi, and G.~Marrocco, ``{RFID Sensors for
  the Monitoring of Body Temperature and Respiratory Function: a Pandemic
  Prospect},'' in \emph{2021 6th International Conference on Smart and
  Sustainable Technologies (SpliTech)}, 2021, pp. 1--5.

\bibitem{Guan20}
W.-J. Guan \emph{et~al.}, ``{Clinical Characteristics of Coronavirus Disease
  2019 in China},'' \emph{New England Journal of Medicine}, vol. 382, February
  2020.

\bibitem{Panunzio21SS}
N.~Panunzio and G.~Marrocco, ``{SECONDSKIN Project: BioIntegrated Wireless
  Sensors for the Epidermal Monitoring and Restoring of Sensorial Injuries},''
  in \emph{2021 IEEE International Conference on RFID Technology and
  Applications (RFID-TA) (IEEE RFID-TA 2021)}, October 2021.

\bibitem{Bianco21RFIDResearch}
G.~M. Bianco, N.~Panunzio, and G.~Marrocco, ``{RFID} research against
  {COVID-19} -- {S}ensorized face masks,'' in \emph{11th IEEE Int. Conf. RFID
  Technol. Appl.}, 2021, to be published.

\bibitem{Massaroni19}
C.~Massaroni, A.~Nicolo, D.~Presti, M.~Sacchetti, S.~Silvestri, and E.~Schena,
  ``{Contact-Based Methods for Measuring Respiratory Rate},'' \emph{Sensors},
  vol.~19, p. 908, February 2019.

\bibitem{Passavanti21}
M.~Passavanti, A.~Argentieri, D.~M. Barbieri, B.~Lou, K.~Wijayaratna, A.~S.
  {Foroutan Mirhosseini}, F.~Wang, S.~Naseri, I.~Qamhia, M.~Tangerås,
  M.~Pelliciari, and C.-H. Ho, ``{The psychological impact of COVID-19 and
  restrictive measures in the world},'' \emph{Journal of Affective Disorders},
  vol. 283, pp. 36--51, 2021.

\bibitem{Daanen20}
H.~Daanen \emph{et~al.}, ``{COVID-19 and thermoregulation-related problems:
  Practical recommendations},'' \emph{Temperature}, August 2020.

\bibitem{Sonner15}
Z.~S. and, ``{The microfluidics of the eccrine sweat gland, including biomarker
  partitioning, transport, and biosensing implications.}''
  \emph{Biomicrofluidics}, vol.~93, p. 031301, 2015.

\bibitem{Buyya19}
R.~Buyya and S.~Srirama, \emph{Fog and edge computing: Principles and
  paradigms}, 2019.

\bibitem{Nordlinger20}
B.~Nordlinger, C.~Villani, and D.~Rus, \emph{Healthcare and artificial
  intelligence}, 2020.

\bibitem{Tarantini2019}
C.~Tarantini, P.~Brouqui, R.~Wilson, K.~Griffiths, P.~Patouraux, and
  P.~Peretti-Watel, ``Healthcare workers' attitudes towards hand-hygiene
  monitoring technology,'' \emph{Journal of Hospital Infection}, vol. 102,
  no.~4, pp. 413--418, 2019.

\bibitem{Lee2009}
J.~Lee and K.~El-Khatib, ``A privacy-enabled architecture for an {RFID-based}
  location monitoring system,'' in \emph{Proc. 7th ACM Int. Symp. Mobility
  Manage. Wireless Access}, Tenerife, Spain, October 2009, pp. 128--131.

\bibitem{Oztekin2010}
A.~Oztekin, F.~Pajouh, D.~Delen, and L.~Swim, ``An {RFID} network design
  methodology for asset tracking in healthcare,'' \emph{Decision Support
  Systems}, vol.~49, no.~1, pp. 100--109, 2010.

\bibitem{Elewe2016}
A.~Elewe, K.~Hasnan, and A.~Nawawi, ``Review of {RFID} optimal tag coverage
  algorithms,'' \emph{ARPN Journal of Engineering and Applied Sciences},
  vol.~11, no.~12, pp. 7706--7711, 2016.

\bibitem{Panunzio21}
N.~{Panunzio}, C.~{Occhiuzzi}, and G.~{Marrocco}, ``{Propagation modeling
  inside the International Space Station for the automatic monitoring of
  Astronauts by means of Epidermal UHF-RFID Sensors},'' \emph{IEEE Journal of
  Radio Frequency Identification}, pp. 1--1, 2021.

\bibitem{VanderTogt2011}
R.~Van~der Togt, P.~Bakker, and M.~Jaspers, ``A framework for performance and
  data quality assessment of radio frequency identification ({RFID}) systems in
  health care settings,'' \emph{Journal of Biomedical Informatics}, vol.~44,
  no.~2, pp. 372--383, 2011.

\bibitem{Chong2015}
A.~Yee-Loong~Chong, M.~Liu, J.~Luo, and O.~Keng-Boon, ``Predicting {RFID}
  adoption in healthcare supply chain from the perspectives of users,''
  \emph{International Journal of Production Economics}, vol. 159, pp. 66--75,
  2015.

\bibitem{Huang18}
C.~Huang, S.~Zhou, J.~Xu, Z.~Niu, R.~Zhang, and S.~Cui, \emph{Energy harvesting
  wireless communications}, 2018.

\bibitem{Georgiadis14}
A.~Georgiadis, ``Energy harvesting for autonomous wireless sensors and
  {RFID}'s,'' in \emph{2014 XXXIth URSI General Assembly and Scientific
  Symposium (URSI GASS)}, 2014, pp. 1--5.

\bibitem{Abdelnour19}
A.~Abdelnour, A.~Hallet, S.~Dkhil, P.~Pierron, D.~Kaddour, and S.~Tedjini,
  ``Energy harvesting based on printed organic photovoltaic cells for {RFID}
  applications,'' in \emph{2019 IEEE International Conference on RFID
  Technology and Applications, RFID-TA 2019}, 2019, pp. 110--112.

\bibitem{Shieh14}
P.~Shieh, N.~Azana, T.~Santos, A.~Martins, N.~Dias, and J.~Xavier, A.L.,
  ``Methodology for choosing piezoelectric devices using piezoelectric energy
  harvesting to feed massive use of {RFID} tags,'' in \emph{2014 IEEE Brasil
  RFID}, 2014, pp. 46--49.

\bibitem{Khemmanee19}
B.~Khemmanee, K.~Tattiwong, and C.~Bunlaksananusorn, ``Experimental study of a
  vibration based energy harvesting system for an {RFID} module,'' in
  \emph{ECTI-CON 2018 - 15th International Conference on Electrical
  Engineering/Electronics, Computer, Telecommunications and Information
  Technology}, 2019, pp. 465--468.

\bibitem{Jauregi17}
I.~Jauregi \emph{et~al.}, ``{UHF RFID} temperature sensor assisted with
  body-heat dissipation energy harvesting,'' \emph{IEEE Sensors Journal},
  vol.~17, no.~5, pp. 1471--1478, 2017.

\bibitem{Li18}
S.~Li, L.~D. Xu, and S.~Zhao, ``{5G Internet of Things: A survey},''
  \emph{Journal of Industrial Information Integration}, vol.~10, pp. 1--9,
  2018.

\end{thebibliography}

\end{document}